\documentclass[aps,showpacs,prd,twocolumn,nofootinbib,nobibnotes,]{revtex4-2}
\usepackage{graphicx}
\usepackage{amssymb}
\usepackage{amsmath}
\usepackage{color}
\usepackage{float}
\usepackage{ulem}
\usepackage{accents}
\usepackage{graphicx}
\usepackage{graphicx}
\usepackage{amsfonts}
\usepackage[colorlinks=true,
pdfstartview=FitV,linkcolor=blue,
citecolor=blue,urlcolor=blue,breaklinks=true]
{hyperref}
\usepackage{array}
\usepackage{float}
\usepackage{placeins}
\usepackage[dvipsnames]{xcolor}
\usepackage{csquotes}
\usepackage{bbold}
\usepackage{units}
\usepackage{enumitem}
\usepackage{dsfont}
\usepackage{upgreek}
\usepackage{makecell}
\newcolumntype{C}[1]{>{\centering\arraybackslash}m{#1}}

\renewcommand{\eqref}[1]{\mbox{Eq.~(\ref{#1})}}

\definecolor{ForestGreen}{rgb}{0.13,0.55,0.13}

\usepackage{fixmath}

\newcommand{\orcid}[1]{\href{https://orcid.org/#1}{\includegraphics[width=10pt]{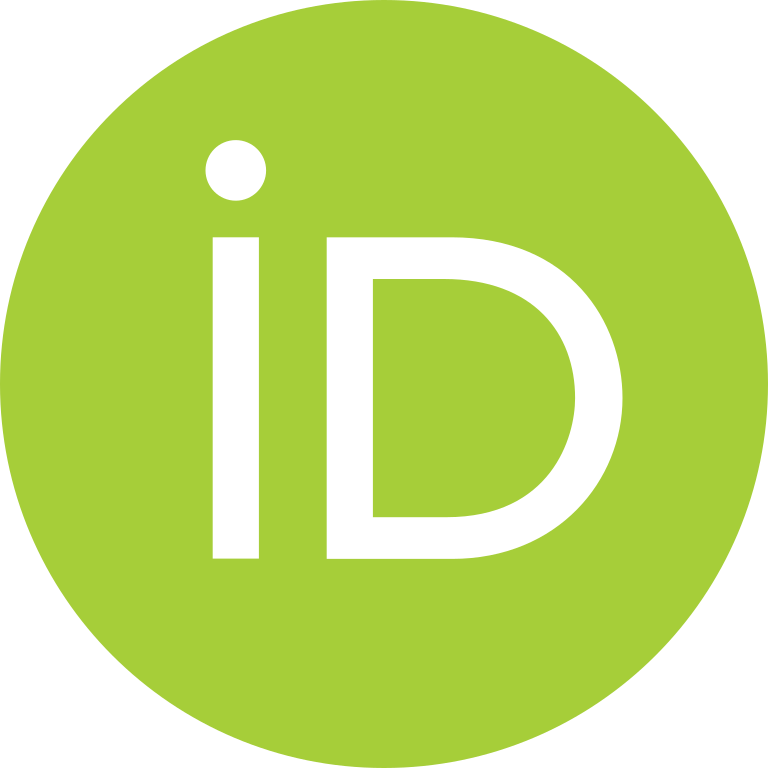}}}

\begin{document}
	
	\title{Anisotropic cold plasma modes in chiral vector Maxwell-Carroll-Field-Jackiw electrodynamics}

\author{Filipe S. Ribeiro\orcid{0000-0003-4142-4304}$^a$}
\email{filipe.ribeiro@discente.ufma.br, filipe99ribeiro@hotmail.com}
	\author{Pedro D. S. Silva\orcid{0000-0001-6215-8186}$^a$}
	\email{pedro.dss@ufma.br, pdiego.10@hotmail.com}
	\author{Manoel M. Ferreira Jr.\orcid{0000-0002-4691-8090}$^b$}
	\email{manojr.ufma@gmail.com, manoel.messias@ufma.br}
		\affiliation{$^a$Programa de P\'{o}s-graduaç\~{a}o em F\'{i}sica, Universidade Federal do Maranh\~{a}o, Campus
		Universit\'{a}rio do Bacanga, S\~{a}o Lu\'is (MA), 65080-805, Brazil}
	\affiliation{$^b$Departamento de F\'{i}sica, Universidade Federal do Maranh\~{a}o, Campus
		Universit\'{a}rio do Bacanga, S\~{a}o Lu\'is (MA), 65080-805, Brazil}

	\begin{abstract}
	In this work, we study the propagation and absorption of plasma waves in the context of the Maxwell-Carroll-Field-Jackiw (MCFJ) electrodynamics with a purely spacelike background playing the role of the anomalous Hall conductivity, concerning the anomalous Hall current. Such a current is also found in an axion field which increases linearly with a space coordinate. The Maxwell equations are rewritten for a cold, uniform, and collisionless fluid plasma model, allowing us to determine the new refractive indices and propagating modes. The analysis begins for propagation along the magnetic axis, examined in the cases of chiral vectors parallel and orthogonal to the magnetic field. Two distinct refractive indices (associated with right-handed circularly polarized [RCP] and left-handed circularly polarized [LCP] waves) are obtained and the associated propagation and absorption zones are determined. The low-frequency regime is discussed and we obtain RCP and LCP helicons. We scrutinize optical effects, such as birefringence and dichroism, and observe rotatory power sign reversion, a property of chiral MCFJ plasmas. We also examine the case of transversal propagation in the direction orthogonal to the magnetic field, providing much more involved results.

\end{abstract}

	\maketitle

\section{Introduction \label{themodel1}}
The propagation properties of electromagnetic waves in a cold magnetized plasma is based on the standard Maxwell equations to describe radio-wave propagation in the ionosphere \cite{refZANGWILL,refJACKSON, chapter-8,Appleton32,refHartree,RATCLIFF2}. The interaction of electromagnetic waves and atmosphere has attracted the attention of researchers over the years, including new investigations on reflection, absorption, and transmission in topical plasma scenarios \cite{Lin-Jing-Guo}. The cold plasma limit is adopted to study the fluid plasma behavior under the action of a constant external magnetic field \cite{Gurnett,STURROK,Stix,Bittencourt,Piel,Boyd}, being defined when the excitation energies are small, so that the thermal and collisional effects can be neglected. In this regime, the ions can be taken as infinitely massive, in such a way that they do not respond to electromagnetic oscillations, especially high-frequency waves. The cold plasma behavior is then described by considering first-order differential equations written for the electron number density $n$ and the electron fluid velocity $\mathbf{u}$, namely,
	\begin{gather}
		\frac{\partial n}{\partial t}+\mathbf{\nabla}\cdot\left(  n\mathbf{u}\right)
		=0,\label{7}\\
		\frac{\partial\mathbf{u}}{\partial t}+\mathbf{u}\cdot\mathbf{\nabla u}%
		=\frac{q}{m}\left(  \mathbf{E}+\mathbf{u}\times\mathbf{B}_{0}\right), \label{7.a}
	\end{gather}
where $\mathbf{B}_{0}$ represents the average magnetic field, and $q$ and $m$ stand for the (electron) charge and mass. The linearized version of the magnetized cold plasmas considers fluctuations around average quantities, $n_{0}$ and $\mathbf{B}_{0}$, which are constant in space and time. Following the usual procedure, assuming $\mathbf{B}%
_{0}=B_{0}\hat{z}$, the corresponding plasma dielectric tensor is
\begin{equation}
\varepsilon_{ij}  (\omega)=\varepsilon_{0}%
\begin{bmatrix}
S & -iD & 0\\
iD & S & 0\\
0 & 0 & P
\end{bmatrix}
,\label{7.1}%
\end{equation}
where $\varepsilon_{0}$ is the vacuum electric permittivity, and
\begin{equation}
S=1-\frac{\omega_{p}^{2}}{\left(  \omega^{2}-\omega_{c}^{2}\right)
},\   D=\frac{\omega_{c}\omega_{p}^{2}}{\omega\left(  \omega^{2}-\omega
	_{c}^{2}\right)  },\  P=1-\frac{\omega_{p}^{2}}{\omega^{2}}, \label{def}
\end{equation}
with $\omega_{p}=n_{0}q^{2}/(m\epsilon_{0})$ and $\omega_{c}=|q| B_{0}/m$ being the plasma and cyclotron frequencies, respectively.

From the Maxwell theory, two distinct refractive indices are obtained for longitudinal propagation to the magnetic field, $\mathbf{k}\parallel\mathbf{B}_{0}$,
\begin{equation}
n_{\pm}=\sqrt{1- \frac{\omega_{p}^{2}} {\omega\left(
		\omega\pm\omega_{c}\right)}} , \label{nusual2}
\end{equation}
which provide right-handed circularly polarized (RCP) and left-handed circularly polarized (LCP) modes \cite{refZANGWILL}. This is the standard result of wave propagation in the usual magnetized cold plasma. The refractive indices (\ref{nusual2}) present the cutoff frequencies $\omega_{\pm}$,
\begin{align}
\omega_{\pm} &=\frac{1}{2}\left(  \sqrt{\omega_{c}^{2}+4\omega_{p}^{2}}\mp\omega_{c}\right), \label{r1}
\end{align}
{defining limits for} the propagation and absorption zones. {As for} propagation orthogonal to the magnetic field, $\mathbf{k}\perp\mathbf{B}_{0}$, $\mathbf{k}=(k_{x},k_{y},0)$, it is found that the corresponding transversal mode, $\delta\mathbf{E}=(0,0,\delta E_{z})$, is associated with the refractive index
	\begin{equation}
	n_{T}=\sqrt{1-\frac{\omega_{p}^{2}}{\omega^{2}}}, \label{nusualort1}
	\end{equation}
	while the extraordinary longitudinal mode, $\delta\mathbf{E}=(\delta E_{x},\delta E_{y},0)$ is related to \cite{Boyd}, 
	\begin{equation}
	n_{O}=\sqrt{\frac{\left(S+D\right)\left(S-D\right)}{S}}, \label{nusualort2}
	\end{equation}
with $P$, $S$, and $D$ given in \eqref{def}. The refractive index $n_{T}$ provides a linearly polarized propagating mode, whereas $n_{O}$, in general, is related to an elliptically polarized mode.

In condensed matter systems, chiral media are endowed with optical activity \cite{TangPRL} stemming from parity-odd models, as bi-isotropic \cite{Sihvola} and bi-anisotropic electrodynamics \cite{Bianiso,Kong,Aladadi,Mahmood,Lorenci,Pedro3}, where circularly polarized waves propagate at distinct phase velocities, yielding birefringence and optical rotation \cite{Fowles}. Such an optical activity is due to anisotropies of the matter structure or can be implied by external fields  (Faraday effect \cite{Bennett, Porter, Shibata}), being measured in terms of rotatory power (RP) \cite{Condon}. Magneto-optical effects constitute a useful tool to investigate new materials, such as topological insulators \cite{Chang1, Urrutia,Urrutia2,Urrutia3,Lakhtakia, Winder, Li, Li1, Tse} and graphene compounds \cite{Crasee}.

In the context of modified electrodynamics, the Maxwell-Carroll-Field-Jackiw (MCFJ) electrodynamics was initially proposed to examine the possibility of \textit{CPT} and Lorentz violation (LV) in free space, establishing severe constraints on the magnitude of the LV coefficients \cite{CFJ}.  This model also represents the \textit{CPT}-odd piece of the \textit{U}(1) gauge sector of the broad Standard Model extension (SME) \cite{Colladay}. The SME has been extensively examined by many authors and in a variety of scenarios, such as in radiative evaluations \cite{CFJ2,CFJ4}, topological defects solutions \cite{CFJ5}, supersymmetry \cite{CFJ6}, Cherenkov radiation \cite{CFJ6A,CFJ6B,CFJ6C}, and classical and quantum aspects \cite{CFJ7}. 
The MCFJ electrodynamics is also relevant due to its connection with the axion Lagrangian \cite{Sekine,Tobar}, 
	\begin{align}
		\mathrm{{\mathcal{L}}}=-\frac{1}{4}F^{\mu\nu}F_{\mu\nu}+\theta (\mathbf{E}\cdot \mathbf{B)},\label{Laxion}
	\end{align}
where $F_{\mu\nu}=\partial_{\mu}A_{\nu}-\partial_{\nu}A_{\mu}$ is the field strength and the axion term, $\theta \, \tilde{F}^{\alpha\beta}F_{\alpha\beta}$, implies
\begin{align}
	\mathrm{{\mathcal{L}}}=-\frac{1}{4}F^{\mu\nu}F_{\mu\nu}  +\frac{1}{4}%
	\epsilon^{\mu\nu\alpha\beta}\left( \partial_{\mu}\theta \right) A_{\nu}F_{\alpha
		\beta}       .\label{Laxion2}
\end{align}
with the dual tensor, $ \tilde{F}^{\mu\nu}=(1/2)\epsilon^{\mu\nu\alpha\beta}F_{\alpha\beta}$. In the case where the axion derivative is a constant vector, $\partial_{\mu}\theta=(k_{AF})_{\mu}$, the Lagrangian (\ref{Laxion2}) recovers the MCFJ one, 
\begin{align}
\mathrm{{\mathcal{L}}}=-\frac{1}{4}G^{\mu\nu}F_{\mu\nu}    + \frac{1}{4}%
\epsilon^{\mu\nu\alpha\beta}\left( k_{AF}\right)_{\mu}A_{\nu}F_{\alpha
	\beta}                         -A_{\mu}J^{\mu}      ,\label{MCFJMATTER}
\end{align}
where $\left(k_{AF}\right)_{\mu}$ is the LV 4-vector background and $G^{\mu\nu}=\frac{1}{2}\chi^{\mu\nu\alpha\beta}F_{\alpha\beta}$ is the continuous matter field strength.\footnote{The 4-rank tensor, $\chi^{\mu\nu\alpha\beta}$, describes the medium constitutive tensor \cite{refPOST}, whose components provide the electric and magnetic responses of the medium. Indeed, the electric permittivity and magnetic permeability tensor components are written as $\epsilon_{ij}\equiv \chi^{0ij0}$ and $\mu^{-1}_{lk}\equiv \frac{1}{4} \epsilon_{ijl}\chi^{ijmn}\epsilon_{mnk}$, respectively. For isotropic polarization and magnetization, it holds $\epsilon_{ij}=\epsilon  \delta_{ij}$ and $\mu^{-1}_{ij}=\mu^{-1} \delta_{ij}$, providing the usual isotropic constitutive relations, $\mathbf{D}=\epsilon\mathbf{E}$, $\mathbf{H}= \mu^{-1}\mathbf{B}$.} Such a Lagrangian provides the modified electrodynamics in matter, described by the inhomogeneous Maxwell equations,
\begin{align}
	\nabla\cdot\mathbf{D}  &=J^{0}-\mathbf{k}_{AF}%
	\cdot\mathbf{B}  ,\label{Coulomb1}\\
	\nabla\times\mathbf{H} -\frac{\partial\mathbf{D}
	}{\partial t} &=\mathbf{J}  -  k_{AF}^{0}    \mathbf{B}+\mathbf{k}_{AF}\times
	\mathbf{E} ,\label{Amp1} 
\end{align}
where $G^{i0}=D^{i}$,  $G^{ij}=-\epsilon_{ijk}H^{k}$,  and $\quad (k_{AF})^{\mu}=(k_{AF}^{0}, \mathbf{k}_{AF})$.
These must be considered
together with the homogeneous Maxwell equations, obtained from the Bianchi identity $\partial_{\mu} \tilde{F}^{\mu\nu}=0$, and suitable constitutive relations.
                                                                                              
The timelike CFJ component, $k_{AF}^{0}$, appears in the modified Amp\`ere's law (\ref{Amp1}) composing the chiral magnetic current,
\begin{equation} 
\mathbf{J}_{B}=k_{AF}^{0}\mathbf{B}, 
\end{equation}
which has been used to investigate electromagnetic properties of matter endowed with the chiral magnetic effect (CME) \cite{Qiu}. The CME \cite{Kharzeev1, Fukushima} consists of a macroscopic linear magnetic current law, ${\bf J}={\sigma}_{B}{\bf B}$, stemming from an asymmetry
between the number density of left- and right-handed chiral fermions. Such an effect has been investigated in a plethora of distinct contexts \cite{Schober, Vilenkin,Maxim, Maxim1, Leite, Dvornikov, Maxim1}, including Weyl semimetals (WSMs) \cite{Burkov}, where the chiral current may be different from the usual CME linear relation when electric and magnetic fields are applied, yielding a current effectively proportional to $B^{2}$, that is,  ${\bf J}={\sigma}({\bf E} \cdot {\bf B}){\bf B}$ \cite{Li,Xiaochun-Huang,Barnes}.

The spacelike vector, $\mathbf{k}_{AF}$, describes an anomalous charge density in Gauss' law (\ref{Coulomb1}) and contributes to the current density,
\begin{equation}
\mathbf{J}_{AH}=\mathbf{k}_{AF} \times\mathbf{E},
\end{equation}
in the Amp\`ere's law (\ref{Amp1}), associated with the anomalous Hall effect (AHE), with $\mathbf{k}_{AF}$ playing the role of anomalous Hall conductivity \cite{Qiu}. The AHE engenders an electric current in the presence of an electric field, due to the separation between the energy-crossing points in momentum space for right-handed and left-handed fermions \cite{ Haldane,Xiao,Huang,Liu}. It has been investigated in distinct contexts, such as noncollinear antiferromagnets \cite{Nakatsuji,Chen}, chiral spin liquids \cite{Machida}, and WSMs \cite{Cote}. Optical effects of a WSM with broken time-reversal and inversion symmetries governed by the axion and the AHE terms were recently examined in WSM systems, with a focus on magneto-optical (Faraday, Kerr, and Voigt) effects \cite{Cote,Trepanier}. The AHE term has also been considered in the propagation of surface plasmon polaritons in WSMs \cite{Gorbar}. Optical effects induced by the current term $\mathbf{k}_{AF} \times\mathbf{E}$ were also examined in the context of the MCFJ electrodynamics in continuous media \cite{Pedroo}. {The anomalous Hall current is also connected with a static axion scenario, $\partial_{t} \theta=0$, with a constant gradient $\mathbf{\nabla}\theta=cte$, as considered to address an axionic Casimir-like effect in Ref. \cite{Brevik2}.}

In a recent investigation \cite{Filipe1}, the chiral effects of the CFJ timelike (pseudoscalar)                                                                    chiral component, $k_{AF}^{0}$, on the electromagnetic modes in magnetized cold plasmas were addressed. The electromagnetic and optical properties of the propagating modes, such as birefringence, absorption, and optical rotation, were discussed, with careful comparisons with the usual cold plasma features allowing the identification of the role played by the chiral factor.

In this work, we study wave propagation in a magnetized cold plasma governed by the Maxwell equations (\ref{Coulomb1}) and (\ref{Amp1}) modified by the AHE current term, $\mathbf{J}_{AH}=\mathbf{k}_{AF} \times\mathbf{E}$, which, using plane-wave ansatz, read
\begin{subequations}
	\label{maxwell-equations-plane-wave-ansatz-1}
	\begin{align}
		i\varepsilon_{ij}{k^i} {{E^j}} +\mathbf{k}_{AF}\cdot \mathbf{{B}}  &=0, \label{4.aac}\\
		i\mathbf{k}\times \mathbf{{B}} +i \mu_{0}\omega \varepsilon_{ij}{k^i} {{E^j}} - \mu_{0}\mathbf{k}_{AF} \times \mathbf{{E}} &=0. \label{4.aab}
	\end{align}
\end{subequations}
We also consider anisotropic constitutive relations (in the electric polarization sector),  
	\begin{equation}
		D^{i}=\varepsilon_{ij}(\omega)E^{j}, \ \ \  B^{i}= \mu_{0} H^{i}. \label{CRAM}
	\end{equation} 
where $\varepsilon_{ij}$ is the cold plasma permittivity (\ref{7.1}) and $\mu_{0}$ is the vacuum permeability. The modified wave equation for the electric field is
	\begin{equation}
	M_{ij} {E}^{j}=0,\label{general.1}				
	\end{equation}
with
\begin{equation}
M_{ij}=n^{2}\delta_{ij}-n_{i}n_{j}-\frac{\varepsilon_{ij}}{\varepsilon_{0}}- \frac{i}{\omega} \epsilon_{ikj}V^{k},
\end{equation} 
written in terms of the refractive index   $\mathbf{n}=\mathbf{k}/\omega$ and with $V^{k}=k_{AF}^{k}/\varepsilon_{0}$ appearing as the redefined components of the chiral vector (which breaks the time-reversal symmetry and preserves space inversion). In this scenario, the wave equation (\ref{general.1}) becomes
	\begin{equation}
	\left[n^{2}\delta_{ij}-n^{i}n^{j}-\frac{{\varepsilon}_{ij}%
	}{\varepsilon_{0}}-i \frac{V^{k}}{\omega} \epsilon_{ikj}\right] {E}^{j}=0,
	\label{EWE2}
	\end{equation}
from which arise the dispersion relations that describe the wave propagation in the medium (by setting $\det M_{ij}=0$).
To obtain the electromagnetic collective modes of a cold plasma modified by the anomalous Hall current-like term, one implements the plasma permittivity tensor (\ref{7.1}) in the wave equation (\ref{EWE2}), yielding the linear homogeneous system
\begin{widetext}
	\begin{equation}
		\begin{bmatrix}
			n^{2}-n_{x}^{2}-S & -n_{x}n_{y}+iD+i\left(V_{z}/\omega\right) & -n_{x}n_{z}-i\left(V_{y}/\omega\right)\\
			-n_{x}n_{y}-iD-i\left(V_{z}/\omega\right) & n^{2}-n_{y}^{2}-S &
			-n_{y}n_{z}+i\left(V_{x}/\omega\right)\\
			-n_{x}n_{z}+i\left(V_{y}/\omega\right) & -n_{y}n_{z}-i\left(V_{x}/\omega\right) & n^{2}-n^{2}_{z}-P
		\end{bmatrix}%
		\begin{bmatrix}
			\delta E_{x}\\
			\delta E_{y}\\
			\delta E_{z}%
		\end{bmatrix}
		=0.  \label{SLmatriz2}
	\end{equation}
\end{widetext}
The refractive indices and associated propagating modes are also obtained, entailing the examination of the optical effects of birefringence and dichroism.  Each scenario is analyzed in the cases of propagation along the magnetic field and orthogonal to the magnetic field, also known as the Faraday and Voigt configurations, respectively \cite{Kotov}.

This paper is outlined as follows. In Sec.~\ref{Wave-propagation-in-purely-time-like-background-case} we obtain the general dispersion relation for a cold magnetized plasma in the presence of the anomalous Hall current, considering the Faraday configuration. In Sec.~\ref{section-IV} we discuss the general properties of the cold plasma modes in the Voigt configuration. The dispersion relations, refractive indices, and optical properties, such as birefringence and absorption, are determined in all cases examined. Finally, we summarize our results in Sec.~\ref{conclusion}.

\section{Wave propagation along the magnetic field axis \label{Wave-propagation-in-purely-time-like-background-case} }

In this section, we analyze the wave propagation along the magnetic field direction, that is, $\mathbf{n}=n\hat{z}$. Then, it holds that

	\begin{widetext}
		\begin{equation}%
		\begin{bmatrix}
		n^{2}-S & +iD+i\left(  \left\vert \mathbf{V}\right\vert /\omega\right)
		\cos\beta & -i\left(  \lvert \mathbf{V}\rvert /\omega\right)\sin\phi\sin\beta\\
		-iD-i\left(  \left\vert \mathbf{V}\right\vert /\omega\right)  \cos\beta & n^{2}-S &
		+i\left(  \left\vert \mathbf{V}\right\vert /\omega\right)  \sin\beta\cos\phi\\
		i\left(  \lvert \mathbf{V}\rvert /\omega\right)\sin\phi\sin\beta & -i\left(  \left\vert \mathbf{V}\right\vert /\omega\right)  \sin\beta\cos\phi & -P
		\end{bmatrix}%
		\begin{bmatrix}
		\delta E_{x}\\
		\delta E_{y}\\
		\delta E_{z}%
		\end{bmatrix}
		=0      \, ,   \label{SLmatriz2.3}
		\end{equation}
	\end{widetext}
where we have used, without loss of generality, the spherical parametrization
\begin{equation}
	\mathbf{V}=\lvert\mathbf{V}\rvert\left(\sin\beta\cos\phi,\sin\beta\sin\phi,\cos\beta\right), \label{bgparamet}
\end{equation}  
with the angle $\beta$ defined between the external magnetic $\mathbf{B}_{0}$ field and the background vector $\mathbf{V}$. Requiring $\mathrm{det}[M_{ij}] =0$ in \eqref{SLmatriz2.3}, the dispersion relations are given by
	\begin{equation}
	2P \left[D ^2-\left(n^2-S\right)^2\right]+\frac{\left\vert \mathbf{V}\right\vert ^{2}}{\omega^{2}}\left(  P+S-n^{2}\right)+\Gamma_{\beta}=0,\label{DR1A}	
	\end{equation}
	where
	\begin{equation}
	\Gamma_{\beta}=\frac{\left\vert \mathbf{V}\right\vert ^{2}}{\omega^{2}}\cos
	(2\beta)\left(  n^{2}+P-S\right)   +4DP \frac{\left\vert \mathbf{V}
		\right\vert}{\omega}  \cos(\beta). \label{DR1A.1}
	\end{equation}
	
We note that the dispersion relation (\ref{DR1A}) depends only on the $\beta$ angle. Thus, we can organize the analysis of the dispersion relation (\ref{DR1A}) by considering two main scenarios: (i) a chiral vector parallel to the magnetic field, and (ii) a chiral vector orthogonal to the magnetic field.

\subsection{\label{section-IIIA}Chiral vector parallel to the magnetic field}

For chiral vector parallel to the magnetic field,  $\mathbf{V}\parallel\mathbf{B}$, one sets $\beta\to 0$ in \eqref{DR1A}, implying
	\begin{equation}
	P \left(\left(n^2-S\right)^2-\left(D+\lvert \mathbf{V}\lvert/\omega\right)^2\right)=0.\label{DRL}	
	\end{equation}
	Longitudinal waves, with $\mathbf{n}\parallel
	\delta\mathbf{E}$ or $\delta\mathbf{E}=(0,0,\delta E_{z})$, may occur when $P=0$, with nonpropagating vibration at the
	plasma frequency, $\omega=\omega_{p}$. 
	
For transverse waves,   $\mathbf{n}\perp\delta\mathbf{E}$ or $\delta\mathbf{E}=(\delta E_{x},\delta E_{y},0)$, the dispersion relation (\ref{DRL}) simplifies to
	\begin{equation}
	\left(n^2-S\right)^2-\left(D+\lvert \mathbf{V}\lvert/\omega\right)^2=0\label{DR1B}
	\end{equation}
which, with the relations (\ref{def}), provides the following refractive indices:
\begin{equation}
\left(  n_{R}\right)  ^{2}=1-\frac{\omega_{p}^{2}}{\omega\left(  \omega-
	\omega_{c}\right)  }- \frac{\lvert \mathbf{V}\lvert}{\omega},\label{SL1}%
\end{equation} 
\begin{equation}
\left(  n_{L }\right)  ^{2}=1-\frac{\omega_{p}^{2}}{\omega\left(  \omega+
	\omega_{c}\right)  }+ \frac{\lvert \mathbf{V}\lvert}{\omega}.\label{SL2}%
\end{equation}

The indices $n_{R}$ and $n_{L}$ may be real or complex in some frequency ranges, enriching their behavior in comparison to the usual cold plasma one. The propagation and absorption zones are modified by the presence of the chiral vector $\mathbf{V}$, as will be shown later.

The propagating modes associated with the refractive indices, given in Eqs. (\ref{SL1}) and (\ref{SL2}), are obtained as the corresponding eigenvectors (with a null eigenvalue) of \eqref{SLmatriz2.3}. The resulting electric fields are the LCP and RCP modes, namely
\begin{align}
n_{L}  \quad &\rightarrow \quad  \ \mathbf{{E}}_{LCP}=\frac{1}{\sqrt{2}}%
\begin{bmatrix}
1 \\ 
+ \. i\\
0
\end{bmatrix}, \label{ELCP2} \\
n_{R} \quad &\rightarrow \quad  \ \mathbf{{E}}_{RCP}=\frac{1}{\sqrt{2}}%
\begin{bmatrix}
1 \\ 
-i \\
0
\end{bmatrix}.\label{ERCP2}
\end{align}

There are two cutoff frequencies for the refractive index $n_{R}$ in (\ref{SL1}),	\begin{equation}
	\omega_{R}^{\pm}= \frac{1}{2}\left(\left(\omega_{c}+\lvert \mathbf{V}\rvert\right) \pm \sqrt{\left(  \omega_{c}-\lvert \mathbf{V}\rvert\right)
		^{2}+4\omega_{p}^{2}}\right).\label{omegaR}
	\end{equation}
While the cutoff frequency $\omega_{R}^{+}$ is always positive, $\omega_{R}^{-}$ is positive only under the condition
	\begin{equation}
	\lvert \mathbf{V}\rvert >\omega_{p}^{2}/\omega_{c},\label{omegaRcond}
	\end{equation}
for which the index $n_{R}$ presents two (positive) roots.
	The refractive index $n_{L}$ (\ref{SL2}) has a single cutoff frequency,
	\begin{align}
	\omega_{L}&= -\frac{1}{2}\omega_{c}-\frac{1}{2}\lvert \mathbf{V}\rvert+\frac{1}{2}\sqrt{\left(  \omega_{c}-\lvert \mathbf{V}\rvert\right)
		^{2}+4\omega_{p}^{2}},\label{omegaL}%
	\end{align}
	which is positive for the condition, 
	\begin{equation}
	\lvert \mathbf{V}\rvert <\omega_{p}^{2}/\omega_{c}.
	\label{omegaLcond}
	\end{equation}

To examine the behavior of the refractive indices (\ref{SL1}) and (\ref{SL2}), we consider two distinct scenarios, following the conditions (\ref{omegaRcond}) and (\ref{omegaLcond}): 
\begin{enumerate}
\item For $\lvert \mathbf{V}\rvert >\omega_{p}^{2}/\omega_{c}$, the index $n_{L}$ presents no positive root (no cutoff), while the index $n_{R}$ has two positive roots.
\item For $\lvert \mathbf{V}\rvert <\omega_{p}^{2}/\omega_{c}$, both indices $n_{L}$ and $n_{R}$ have one positive cutoff.
\end{enumerate}

\subsubsection{About the index $n_{R}$ \label{secNR}}

\begin{figure}[h]
	\centering
	\includegraphics[scale=0.72]{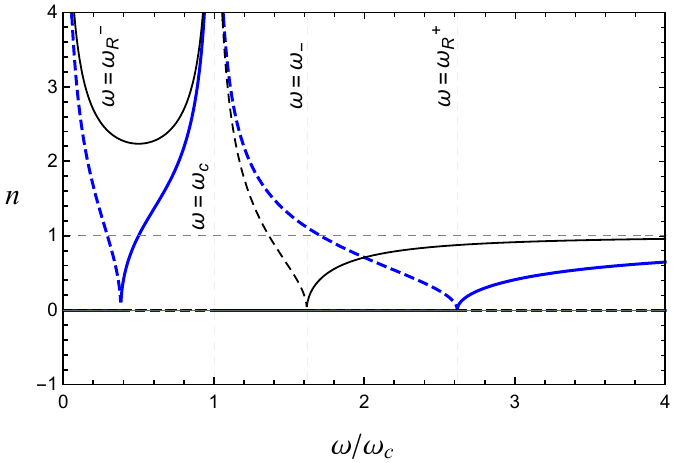}  \caption{Index of refraction 
			$n_{R}$, under the condition (\ref{omegaRcond}). The dashed blue (black) line corresponds to the imaginary piece of $n_{R}$ ($n_{-}$), while the solid blue (black) line represents the real piece of $n_{R}$ ($n_{-}$).  Note that the chiral factor opens a new lossy window near the origin and enlarges the second absorption zone. Here, $\omega_{c}=\omega_{p}$, $\lvert \mathbf{V}\rvert=2\omega_{p}$, and $\omega_{c}=1$~$\mathrm{rad}$~s$^{-1}$. The (dashed and solid)
			blue lines representing $n_{R}$ is thicker than the black line for $n_{-}$.}
	\label{nRfig}%
\end{figure}

The general behavior of the index $n_{R}$ is represented in Fig.~\ref{nRfig} in terms of the dimensionless parameter $\omega/\omega_{c}$ and under the condition (\ref{omegaRcond}), as detailed below.

\begin{enumerate}
	[label=(\roman*)]
	
	\item For $\omega\rightarrow0$, $n_{R}\rightarrow+i\infty$, which is complex and divergent in this limit, differing from the behavior of the usual magnetized plasma index $n_{-}$, which provides $n\rightarrow\infty$ near the origin.
	
	\item For $0<\omega<\omega_{R}^{-}$, an absorption zone appears, where $\mathrm{Im}\left[n_{R}\right]\neq 0$. This characteristic does not manifest in the usual cold plasma index $n_{-}$, which is real and positive in this range. See the black line in this frequency zone in Fig.~\ref{nRfig}.
	
	\item For $\omega_{R}^{-}<\omega<\omega_{c}$, $n_{R}$ is real, with $\mathrm{Re}\left[n_{R}\right]>0$, revealing a propagation zone.
	
	\item For $\omega\rightarrow\omega _{c}$, $n_{R}\rightarrow\infty$,
	 a resonance at the cyclotron frequency occurs.
	
	\item For $\omega_{c}<\omega<\omega_{R}^{+}$, there is an absorption zone, where $n_{R}$ is imaginary, $\mathrm{Im}\left[n_{R}\right]\neq0$. Such an absorption zone is larger than the usual zone shown by the black-dashed line in Fig.~\ref{nRfig}, since $\omega_{R}^{+}>\omega_{-}$.
	
	\item For $\omega>\omega_{R}^{+}$, the index $n_{R}$ is real and positive, yielding an attenuation-free propagating zone, and recovering $n_{R} \rightarrow1$ in the high-frequency limit.
	
\end{enumerate}

\begin{figure}[h]
	\centering
	\includegraphics[scale=0.72]{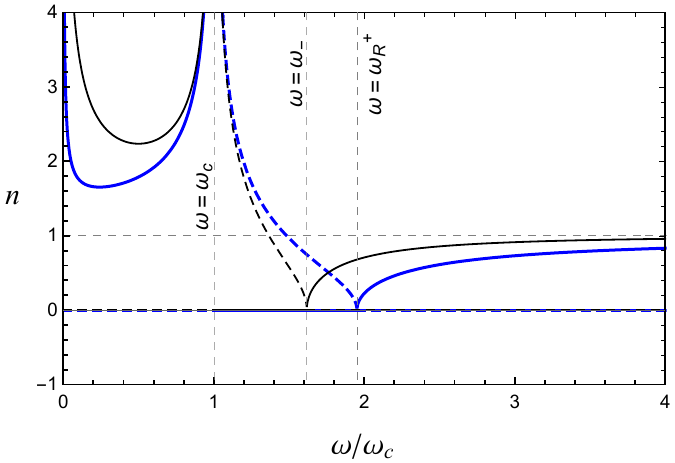}  \caption{Index of refraction 
			$n_{R}$, under the condition (\ref{omegaLcond}). The dashed blue (black) line corresponds to the imaginary piece of $n_{R}$ ($n_{-}$), while the solid blue (black) line represents the real piece of $n_{R}$ ($n_{-}$). The chiral factor increases the absorption zone near resonance. Here, $\omega_{c}=\omega_{p}$, $\lvert \mathbf{V}\rvert=0.5\omega_{p}$, and $\omega_{c}=1$~$\mathrm{rad}$~s$^{-1}$. The (dashed and solid)
			blue lines representing $n_{R}$ is thicker than the black line for $n_{-}$.}
	\label{nRfig2}%
\end{figure}

On the other hand, under condition (\ref{omegaLcond}), $\omega_{R}^{-}<0$, so there is only one cutoff frequency and a single absorption zone, defined for $\omega_{c}<\omega<\omega_{R}^{+}$. The first absorption zone is replaced by a propagation region, now defined for $0<\omega<\omega_{c}$, similar to the usual case. For $\omega>\omega_{c}$, the behavior is similar to that pointed out in items $(v)$ and $(vi)$ above. This scenario for $n_{R}$ is illustrated in Fig.~\ref{nRfig2}.


\subsubsection{About the index $n_{L}$ \label{secNL}}

The index $n_{L}$, given in \eqref{SL2}, has no positive root under the condition (\ref{omegaRcond}) (no cutoff frequency) and one cutoff frequency under the condition (\ref{omegaLcond}). Its features are described below.

\begin{figure}[h]
	\centering
	\includegraphics[scale=0.72]{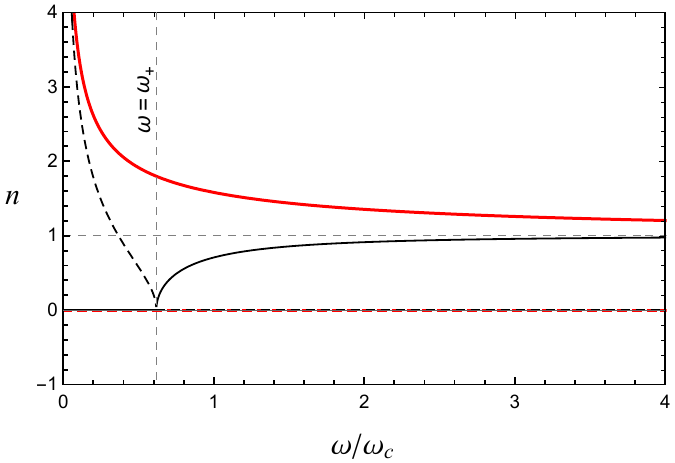}
	\caption{Refractive index $n_{L}$ (red lines) for condition (\ref{omegaRcond}) and index $n_{+}$ (black lines) of \eqref{nusual2}. The dashed (solid) lines correspond to the imaginary (real) pieces of $n_{L}$ and $n_{+}$. The chiral factor suppresses the absorption window. Here, $\omega_{c}=\omega_{p}$, $\lvert \mathbf{V}\rvert=2\omega_{p}$, and $\omega_{c}=1$~$\mathrm{rad}$~s$^{-1}$. The (dashed and solid)
		red line representing $n_{L}$ is thicker than the black line for $n_{+}$.}
	\label{nLfig}	
\end{figure}

\begin{figure}[h]		
	\centering
	\includegraphics[scale=0.72]{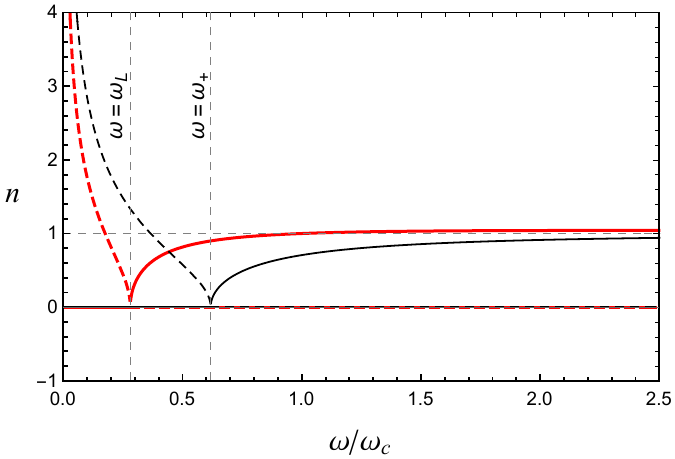}
	\caption{Refractive index $n_{L}$ (red lines) for condition (\ref{omegaLcond}) and refractive index $n_{+}$ (black lines) of \eqref{nusual2}. The dashed (solid) lines correspond to the imaginary (real) pieces of $n_{L}$ and $n_{+}$. The chiral factor narrows the absorption zone and enhances the attenuation-free propagation regime. Here, $\omega_{c}=\omega_{p}$, $\lvert \mathbf{V}\rvert=0.5\omega_{p}$, and $\omega_{c}=1$~$\mathrm{rad}$~s$^{-1}$. The (dashed and solid)
		red lines representing $n_{L}$ is thicker than the black line for $n_{+}$.}
	\label{nLfig2}
\end{figure}

\begin{enumerate}
	[label=(\roman*)]
	
	\item For $\omega\rightarrow0$, under the condition (\ref{omegaRcond}), the presence of $\lvert\mathbf{V}\rvert$ renders the refractive index real and positively divergent at the origin, $n_{L}\rightarrow+\infty$,  differing from the usual index $n_{+}$ (\ref{nusual2}), which is complex and divergent, $\mathrm{Im}[n_{+}]\rightarrow\infty$, at the origin. This behavior is similar to what is observed in the index $n_{L}$ of the chiral MCFJ model examined\footnote{See Eq. (48) and Fig. 4 of Ref.~\cite{Filipe1}.} in Ref.~\cite{Filipe1}. 
		
	\item For $\omega>0$, the index $n_{L}$ is always real, $\mathrm{Im}[n_{L}]=0$. Thus, wave propagation occurs for any frequency. The real and imaginary parts of $n_{L}$ are represented in Fig.~\ref{nLfig}.
	
\item Under the condition (\ref{omegaLcond}), $n_{L}$ presents a cutoff frequency, $\omega_{L}$, given by \eqref{omegaL}. Therefore, an absorption zone appears for $0<\omega<\omega_{L}$, where $\mathrm{Im}\left[n_{L}\right]\neq 0$, as depicted in Fig.~\ref{nLfig2}.
 
\end{enumerate}

We observe that the indices $n_{R}$ and $n_{L}$ are always positive, implying the nonexistence of negative refraction, a phenomenon that was reported in the context of the MCFJ cold plasmas in the presence of the chiral timelike background factor.
 	
The results of this subsecton may be compared with the case of wave propagation along the magnetic field with the MCFJ timelike background, examined in Sec.~IV of Ref~\cite{Filipe1}, whose scenario was richer due to the attainment of four distinct indices and negative refraction, which typically occurs in bi-isotropic media as well \cite{Guo,Gao}. In the present case, however, there are only two positive indices (no negative refraction).\footnote{Note that negative indices may also exist in the case where one takes the negative roots of Eqs.~(\ref{SL1}) and (\ref{SL2}), that is, $-n_{L}$ and $-n_{R}$, which correspond to the exact mirror image of the positive indices (in relation to the frequency axis).} Nevertheless, regarding the propagation and absorption properties, the present chiral-vector case is more involved since two absorption zones are opened under the condition (\ref{omegaRcond}), while in the analogous situation of Ref.~\cite{Filipe1} only one absorption zone was reported.

\subsubsection{Dispersion relation behavior}
\begin{figure}[h]
	\centering
	\includegraphics[scale=0.5]{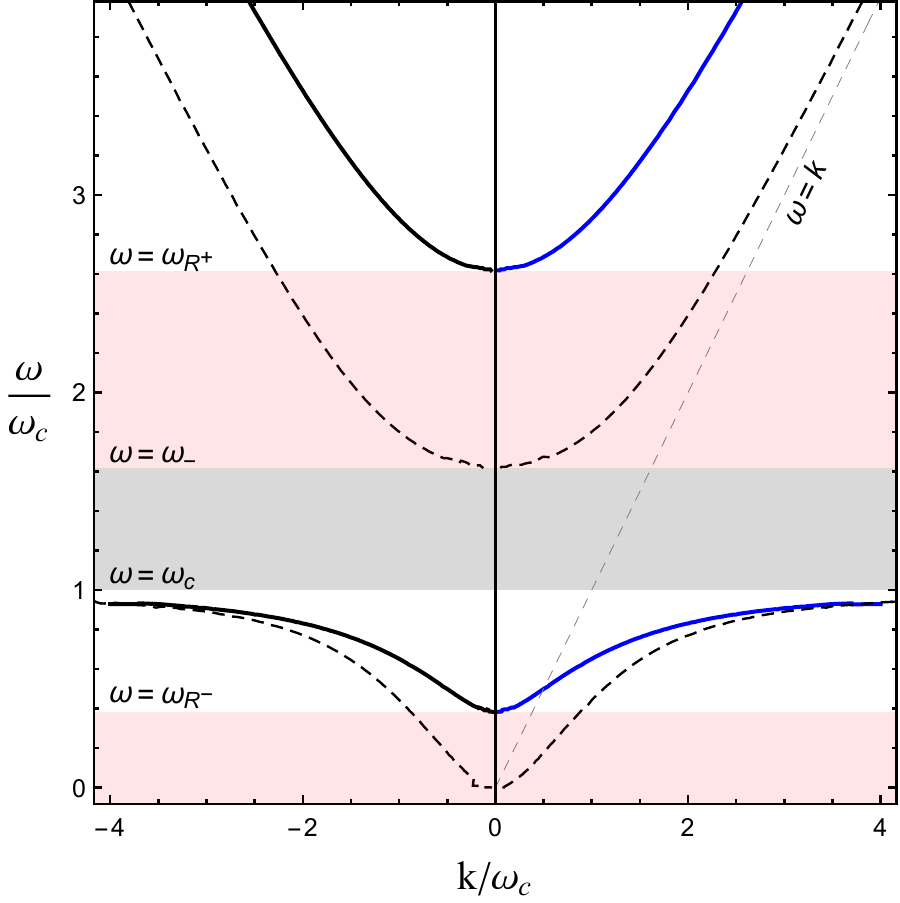}  \caption{Dispersion relations related to the refractive indices $n_{R}$ (solid blue line) and $-n_{R}$ (solid black line), under the  condition (\ref{omegaRcond}). The dashed black line represents the indices of the standard case, $\pm n_{-}$. The highlighted area in red indicates the enlargement of the absorption zone for $n_{R}$ in comparison to the gray absorption zone of the usual indices ($\pm n_{-}$). Here we use $\omega_{c}=\omega_{p}$ and $\lvert \mathbf{V}\rvert=2\omega_{p}$, with $\omega_{c}=1$~$\mathrm{rad}$~s$^{-1}$.}
	\label{oxk_nr_1}%
\end{figure}

\begin{figure}[h]
	\centering
	\includegraphics[scale=0.5]{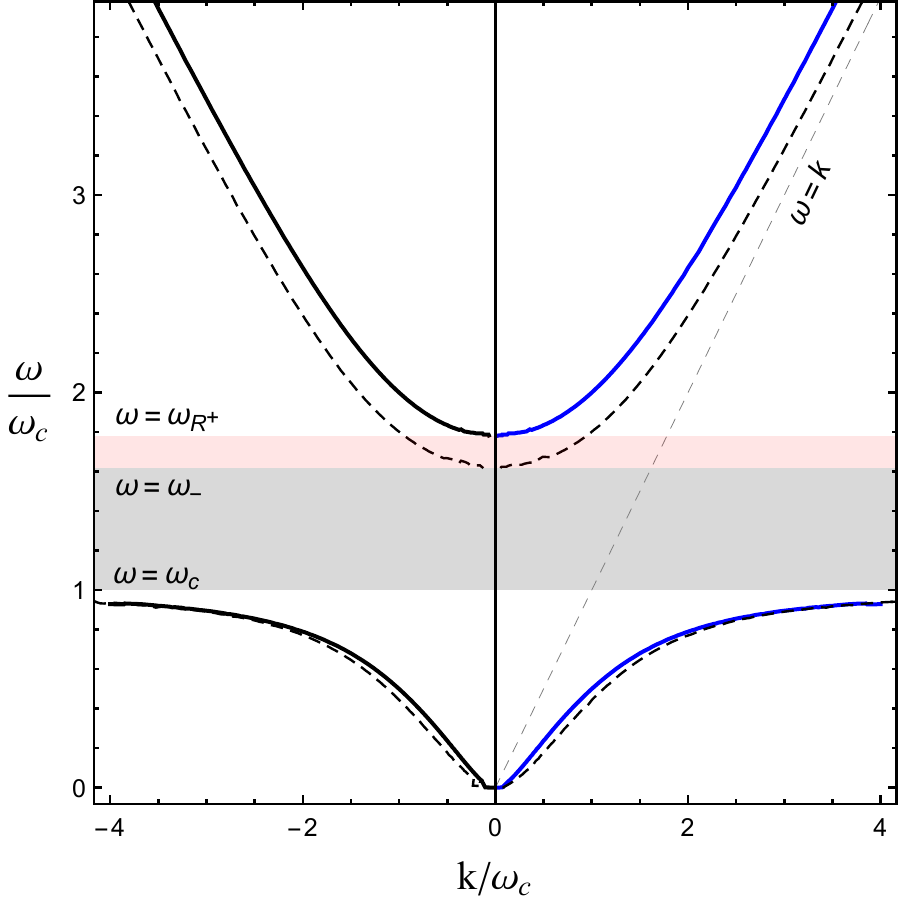}  \caption{Plot of the dispersion relations related to the refractive indices $n_{R}$ (solid blue line) and $-n_{R}$ (solid black line), under the condition (\ref{omegaLcond}). The dashed black line corresponds to the indices of the usual case ($\pm n_{-}$). The highlighted area in red indicates the absorption zone amplification for $n_{R}$ in comparison with the one of $\pm n_{-}$. Here we use $\omega_{c}=\omega_{p}$ and $\lvert \mathbf{V}\rvert=0.5\omega_{p}$, with $\omega_{c}=1$~$\mathrm{rad}$~s$^{-1}$.}
	\label{oxk_nr_2}%
\end{figure}

\begin{figure}[h]
	\centering
	\includegraphics[scale=0.5]{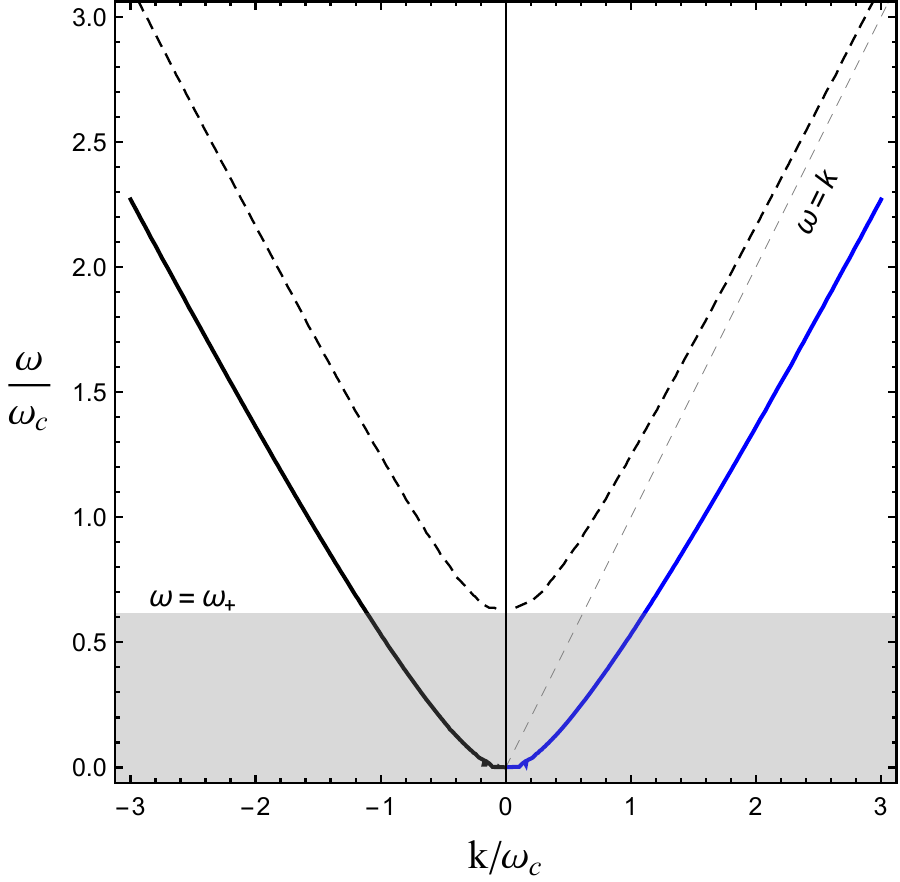}  \caption{Dispersion relations associated with the refractive indices $n_{L}$ (solid blue line) and $-n_{L}$ (solid black line), under the condition (\ref{omegaRcond}). The dashed black line corresponds to the indices of the usual case ($\pm n_{+}$). The highlighted area in gray indicates the absorption zone for the indices $\pm n_{+}$, where one notices the absence of absorption for $n_{L}$. Here we use $\omega_{c}=\omega_{p}$ and $\lvert \mathbf{V}\rvert=2\omega_{p}$, with $\omega_{c}=1$~$\mathrm{rad}$~s$^{-1}$.}
	\label{oxk_nl_1}%
\end{figure}

\begin{figure}[h]
	\centering
	\includegraphics[scale=0.5]{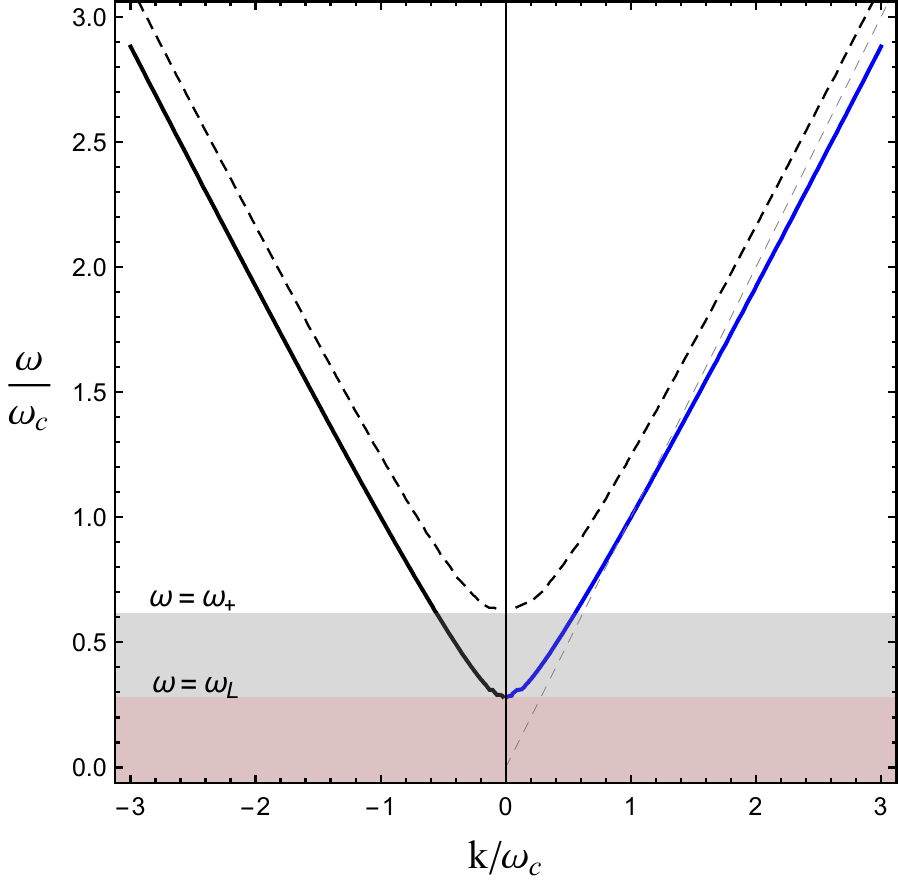}  \caption{Dispersion relations associated with the refractive indices $n_{L}$ (solid blue line) and $-n_{L}$ (solid black line), under the  condition (\ref{omegaLcond}). The dashed black line corresponds to the indices of the usual case ($\pm n_{+}$). The highlighted area in red indicates the absorption zone for $n_{L}$, narrowed in comparison to the absorption zone associated with $\pm n_{+}$. Here we use $\omega_{c}=\omega_{p}$ and $\lvert \mathbf{V}\rvert=0.5\omega_{p}$, with $\omega_{c}=1$~$\mathrm{rad}$~s$^{-1}$.}
	\label{oxk_nl_2}%
\end{figure}

The behavior of the dispersion relations can be visualized in plots of $\omega \times k $. In this subsection, the dispersion relations associated with the circular modes, connected to the indices $n_{R}$ and $n_{L}$, are presented in dimensionless plots of  $\left(\omega/\omega_{c}\right) \times \left(k/\omega_{c}\right)$.
The dispersion relation associated with $\pm n_{R}$, under the condition (\ref{omegaRcond}), is depicted in Fig.~\ref{oxk_nr_1}. The propagation occurs for $ \omega_{R}^{-}<\omega<\omega_{c}$ and $\omega>\omega_{R}^{+}$, while two absorption windows appear:  $\omega_{c}<\omega<\omega_{R}^{+}$ and $0<\omega<\omega_{R}^{-}$. On the other hand, the behavior for the condition (\ref{omegaLcond}) is shown in Fig.~\ref{oxk_nr_2}, where the propagation occurs for $0<\omega<\omega_{c}$ and $\omega>\omega_{R}^{+}$ and the absorption happens for $\omega_{c}<\omega<\omega_{R}^{+}$. Figures~\ref{oxk_nr_1} and \ref{oxk_nr_2} illustrate enhanced absorption zones.
Figure \ref{oxk_nl_1} illustrates the dispersion relations associated with $\pm n_{L}$ for the condition (\ref{omegaRcond}), where the propagation occurs for $\omega>0$, being compatible with the absence of an absorption zone; see Fig.~\ref{nLfig}. The dispersion relation for the condition (\ref{omegaLcond}) is depicted in Fig.~\ref{oxk_nl_2}, where there is an unusual absorption zone in $0<\omega<\omega_{L}$, while the propagation appears for $\omega>\omega_{L}$. In these two latter cases, the absorption zone is reduced in comparison to the zone of the usual indices, $\pm n_{+}$,  $0<\omega<\omega_{+}$.

An interesting point is that the refractive indices $n_{R,L}$, represented by blue lines in Figs. \ref{oxk_nr_1}--\ref{oxk_nl_2}, do not become negative under the influence of the chiral vector $\mathbf{V}$, allowing the plots of $\omega \times k$  to remain centered at $k=0$. This is not the case for cold plasmas under the timelike CFJ electrodynamics \cite{Filipe1}, where the scalar chiral parameter $V_{0}$ induces zones of negative refraction (negative refractive indices), decentralizing the curves of $\omega \times k$ (see Ref.~\cite{Filipe1}).

\subsubsection{\label{section-helicons}Low-frequency helicon modes}
There are low-frequency plasma modes that propagate along the magnetic field axis, called helicons. In a usual magnetized plasma, there exist only RCP helicon modes,\footnote{See Chapter 9 of Ref.~\cite{Bittencourt} and Chapter 8 of Ref.~\cite{chapter-8} for basic details.} for which the refractive index (\ref{nusual2}) yields 
\begin{align}
	n_{-} &= \omega_{p} \sqrt{\frac{1}{\omega \omega_{c}}} \label{helicons-12}
\end{align}
in the low-frequency regime,
\begin{align}
	\omega\ll \omega_{p}, \quad \omega_{c}\ll\omega_{p}, \quad  \omega\ll \omega_{c}. \label{helicon-frequency-regime}
\end{align}

Considering the circular electromagnetic modes associated with the indices (\ref{SL1}) and (\ref{SL2}), the corresponding helicons indices are 
\begin{subequations}
\label{helicons-Faraday-configuration-with-background-parallel-1}
	\begin{align}
		\bar{n}_{R}&=\sqrt{\frac{\omega_{p}^{2}}{\omega \omega_{c} }-\frac{ \lvert\mathbf{V}\rvert}{\omega}}, \label{helicons-15} \\
		\bar{n}_{L} &=\sqrt{\frac{ \lvert\mathbf{V}\rvert}{\omega}-\frac{\omega_{p}^{2}}{\omega \omega_{c} }} , \label{helicons-16}
	\end{align}
\end{subequations}
	where we have used the ``bar'' notation to indicate the helicons quantities. Due to the chiral vector $\lvert\mathbf{V}\rvert$, one obtains expressions for both RCP and LCP helicons. As $\bar{n}_{R}^2=-\bar{n}_{L}^2$, one of these indices is imaginary when the other is real. In fact, $\bar{n}_{R}$ becomes imaginary for $\lvert\mathbf{V}\rvert>\omega_{p}^{2}/\omega_{c}$, while  $\bar{n}_{L}$ is imaginary for $\lvert\mathbf{V}\rvert<\omega_{p}^{2}/\omega_{c}$. This means that RCP and LCP helicons do not propagate simultaneously. Indeed, only one of the modes in \eqref{helicons-Faraday-configuration-with-background-parallel-1} can propagate for each value of $\lvert\mathbf{V}\rvert$ adopted. In this context, note that the usual cold plasma helicon mode is recovered in the limit $\lvert\mathbf{V}\rvert \rightarrow 0$, for which the helicon index $\bar{n}_{R}$ yields the usual result of \eqref{helicons-12}, while $\bar{n}_{L}$ becomes purely imaginary, indicating the absence of propagation.

\subsubsection{Rotatory power \label{secRP}}

 Chiral media possess optical activity, usually described in terms of the rotation of the polarization (birefringence) that takes place when RCP and LCP modes propagate at different phase velocities. Such a rotation is measured in terms of the rotatory power,  which is useful for performing an optical characterization of multiple systems, such as crystals \cite{Dimitriu, Birefringence1},  organic compounds \cite{Barron2, Xing-Liu}, graphene phenomena in the terahertz band \cite{Poumirol}, the gas of fast-spinning molecules \cite{Tutunnikov}, chiral metamaterials \cite{Woo, Zhang, Mun}, and chiral semimetals \cite{Pesin, Dey-Nandy}, and in the determination of the rotation direction of pulsars \cite{Gueroult2}. The RP may be dispersive (depend on the frequency) \cite{Newnham, Tschugaeff, Tischler}. The RP is defined as
	\begin{equation}
		\delta=-\frac{\omega}{2}\left(\mathrm{Re}[n_{L}]-\mathrm{Re}[n_{R}]\right), 
	\end{equation}
where $n_{L}$ and $n_{R}$ are the refractive indices for different circular polarizations. For the case where the background vector is parallel to the magnetic field (see Sec.~\ref{section-IIIA}), the refractive indices, given by Eqs.~(\ref{SL1}) and (\ref{SL2}), yield
	\begin{equation}
		\delta=-\frac{\omega}{2}\mathrm{Re}\left[\sqrt{R_{+}}-\sqrt{R_{-}}\right], \label{rptl}
	\end{equation}
where $R_{\pm}$ is given by
	\begin{equation}
		R_{\pm}= 1-\frac{\omega_{p}^{2}}{\omega\left(  \omega\pm
			\omega_{c}\right)  }\pm \frac{\lvert \mathbf{V}\lvert}{\omega}.
		\label{N+-}
	\end{equation}
	
\begin{figure}[h]
	\centering
	\includegraphics[scale=0.7]{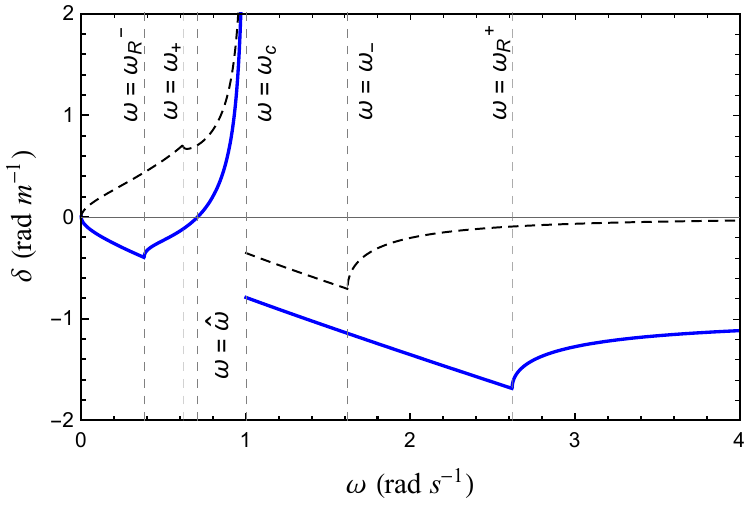}
	\caption{The solid blue line represents the RP (\ref{rptl}) defined by the refractive index $n_{L}$ and $n_{R}$ for the condition (\ref{omegaRcond}). The dashed black line corresponds to the usual RP for a conventional cold plasma. The chiral factor determines the RP sign reversion at $\omega<\hat{\omega}$ and a constant asymptotic value, $-|\mathbf{V}|/2$, for high frequencies. Here we use $\omega_{c}=\omega_{p}$, $\lvert \mathbf{V}\rvert=2\omega_{p}$, and $\omega_{c}=1$~$\mathrm{rad}$~s$^{-1}$. }
	\label{rptl1}
\end{figure}

\begin{figure}[h]
	
	\centering
	\includegraphics[scale=0.7]{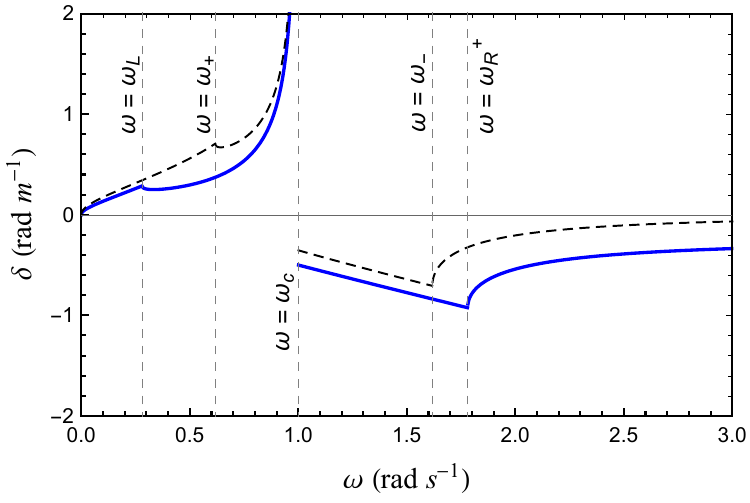}
	\caption{Solid blue lines: RP (\ref{rptl}) associated with the refractive indices $n_{L}$ and $n_{R}$ for the condition (\ref{omegaLcond}). The dashed line represents the usual RP for a usual cold plasma. Small deviations of the RP are originated from the chiral factor. The main difference in comparison with the usual case is the non-null asymptotical value, $-|\mathbf{V}|/2$, for high frequencies. Here we use $\omega_{c}=\omega_{p}$, $\lvert \mathbf{V}\rvert=0.5\omega_{p}$, and $\omega_{c}=1$~$\mathrm{rad}$~s$^{-1}$.}
	\label{rpsl2}
\end{figure}

The behavior of the RP (\ref{rptl}) is depicted in Fig.~\ref{rptl1}, being negative for the interval $0<\omega<\hat{\omega}$ and positive for $\hat{\omega}<\omega<\omega_{c}$.  For $\omega>\omega_{c}$, the RP is always negative, exhibiting a sharp behavior at $\omega=\omega_{R}$, the point at which the real piece of $n_{R}$ assumes nonzero values again (see Fig.~\ref{nRfig}). The frequency $\hat{\omega}$,
\begin{equation}
		\hat{\omega}=\sqrt{\omega_{c}\left(\omega_{c}-\omega_{p}^{2}/\lvert \mathbf{V}\rvert\right)},
		\label{FreqRPR}
	\end{equation}
obtained from the Eqs.~(\ref{rptl}) and (\ref{N+-}), indicates where the RP changes sign, as shown in Fig.~\ref{rptl1}. Note that $\hat{\omega}$ is real only for the condition (\ref{omegaRcond}). Hence the RP reversion only occurs in this case, as confirmed in Fig.~\ref{rptl1}. In fact, under the condition (\ref{omegaLcond}), the corresponding RP depicted in Fig.~\ref{rpsl2} is not endowed with sign reversal,  a behavior analog to the
standard RP in plasmas. 

It is worth remarking that the RP reversion is observed in graphene systems \cite{Poumirol}, Weyl metals and semimetals with low electron density with chiral conductivity \cite{Pesin,Dey-Nandy},  and bi-isotropic dielectrics with magnetic chiral conductivity \cite{PedroPRB}. Such a reversion does not occur in conventional cold plasma, but it takes place in rotating plasmas \cite{Gueroult} and in the MCFJ plasma with a chiral pseudoscalar factor \cite{Filipe1}. Therefore, RP reversion may be considered a signature of chiral MCFJ nonrotating cold plasmas.

Furthermore, it is important to pay attention to the intervals $0<\omega<\omega_{R}^{-}$ and $\omega_{c}<\omega<\omega_{R}^{+}$, where the refractive index $n_{R}$ is imaginary and the RP receives contribution only from the index $n_{L}$,  exhibiting an approximately linearly increasing magnitude, the opposite of the $n_{L}$ profile. For $\omega>\omega_{R}^{+}$, the modes associated with the $n_{L}$ and $n_{R}$ propagate and contribute to the RP, whose magnitude diminishes monotonically with $\omega$, tending to the asymptotic value, $-|{\bf{V}}|/2$, as shown in Fig.~\ref{rptl1}. In fact, in the high-frequency limit, $\omega>>\omega_{p},\omega_{c} $, the refractive indices $n_{L}$ and $n_{R}$ provide (at first order)
			\begin{align}
			n_{L,R} \simeq \,1 \pm \frac{\lvert\mathbf{V}\rvert}{2\omega}, 
			\end{align}
so the RP asymptotic value is
			\begin{equation}
			\delta \simeq-\frac{\lvert\mathbf{V}\rvert}{2},\label{faradayTIMELIKE}
			\end{equation}
		a result that holds even in the absence of the magnetic field. This asymptotic limit differs from the behavior of a cold usual plasma, whose RP decays as $1/\omega^2$ for high frequencies, tending to zero for $\omega>>\omega_{p},\omega_{c} $. See the dashed line in Fig.~\ref{rptl1}.                          

\subsubsection{Dichroism coefficients \label{secDC} }

\begin{figure}[h]	
	\centering
	\includegraphics[scale=0.65]{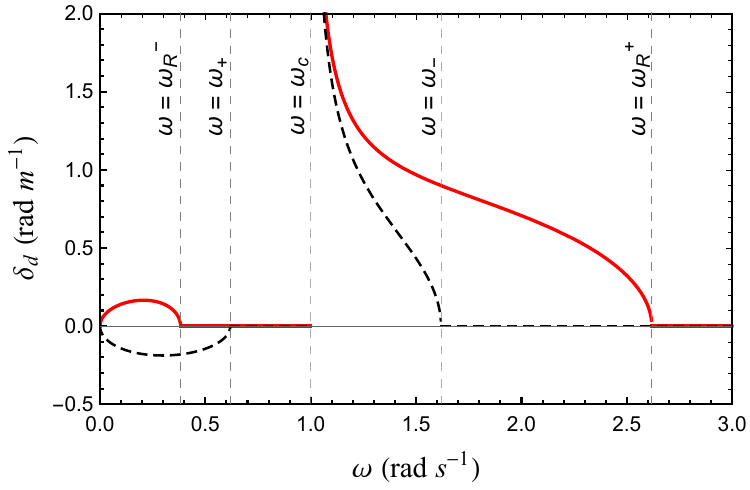}
	\caption{Dichroism coefficient of \eqref{dicrosl} (red solid lines) associated with $n_{L}$ and $n_{R}$, under the condition (\ref{omegaRcond}). The black dashed line represents the dichroism coefficient of a usual cold plasma. Here, $\omega_{c}=\omega_{p}$, $\lvert \mathbf{V}\rvert=2\omega_{p}$, and $\omega_{c}=1$~$\mathrm{rad}$~s$^{-1}$.}
	\label{dicroSL}
\end{figure}

Absorption occurs in the zones where the indices are complex. When circularly polarized modes undergo absorption at different degrees, dichroism takes place,  working as another parameter for optical characterization.  It could be used to distinguish between Dirac and Weyl semimetals \cite{Hosur}, perform enantiomeric discrimination \cite{Nieto-Vesperinas,Tang}, and develop graphene-based devices at terahertz frequencies \cite{Amin}.
Dichroism for LCP and RCP waves is expressed in terms of the coefficient
\begin{equation}
	\delta_{d} =-\frac{\omega }{2}\left( \mathrm{Im}[n_{L}]-\mathrm{Im}[n_{R}]\right).  \label{dicro_eqtl}
\end{equation}
For the condition (\ref{omegaRcond}), $n_{R}$ has non-null imaginary parts in the intervals $0<\omega<\omega_{R}^{-}$ and $\omega_{c}<\omega<\omega_{R}^{+}$, while $n_{L}$ is always real (for $\omega>0$). In this case, the dichroism coefficient is written as
	\begin{equation}
		\delta_{d}=\begin{cases}
			\text{$\frac{\omega}{2}\sqrt{R_{-}},$} &\quad\text{for $0<\omega<\omega_{R}^{-},$}\\
			\text{$0,$} &\quad\text{for $\omega_{R}^{-}<\omega<\omega_{c},$}\\
			\text{$\frac{\omega}{2}\sqrt{R_{-}},$} &\quad\text{for $\omega_{c}<\omega<\omega_{R}^{+},$}\\
			\text{$0,$} &\quad\text{for $\omega>\omega_{R}^{+},$}
		\end{cases}\label{dicrosl}
	\end{equation}
with $R_{-}$ of \eqref{N+-}. Such a coefficient is depicted in Fig.~\ref{dicroSL}.

\begin{figure}[h]	
	\centering
	\includegraphics[scale=0.65]{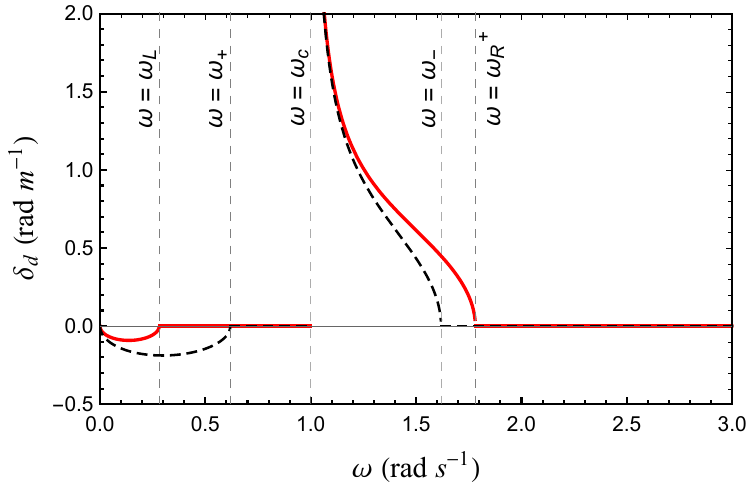}
	\caption{Dichroism coefficient of \eqref{dicro_TL2.3} (solid red lines) associated with $n_{L}$ and $n_{R}$, under the condition (\ref{omegaLcond}). The dashed line represents the dichroism coefficient for a conventional cold plasma. Here, we set $\omega_{c}=\omega_{p}$, $\lvert \mathbf{V}\rvert=0.5\omega_{p}$, and $\omega_{c}=1$~$\mathrm{rad}$~s$^{-1}$.}
	\label{dicroSL2}
\end{figure}

Considering the condition (\ref{omegaLcond}), both $n_{R}$ and $n_{L}$ have non-null imaginary parts in the intervals $\omega_{c}<\omega<\omega_{R}^{+}$ and $0<\omega<\omega_{L}$, respectively, in which the dichroism coefficient is non-null,
	\begin{equation}
		\delta_{d}=\begin{cases}
			\text{$-\frac{\omega}{2} \sqrt{R_{+}}$}, &\quad\text{for $0<\omega<\omega_{L}$},\\
			\text{$+\frac{\omega}{2} \sqrt{R_{-}}$}, &\quad\text{for $\omega_{c}<\omega<\omega_{R}^{+}$}.\\
		\end{cases}     \label{dicro_TL2.3}
	\end{equation}
The general behavior of (\ref{dicro_TL2.3}) is exhibited in Fig.~\ref{dicroSL2}.

\subsection{\label{BGVORT}Chiral vector orthogonal to the magnetic field}

In the scenario where the background vector is 	orthogonal to the magnetic field, $\beta\to\pi/2$, and the dispersion relation (\ref{DR1A}) reads
	\begin{equation}
		P \left[D ^2-\left(n^2-S\right)^2\right] +\left(\lvert \mathbf{V}\rvert^2/\omega^2\right) \left(S-n^2\right)=0,\label{DR2}	
	\end{equation}	
yielding two refractive indices:
	\begin{equation}
		\left(n_{A}\right)^2=S-\frac{\lvert \mathbf{V}\rvert^2}{2 P \omega ^2}+\frac{1}{ P}\sqrt{ P^2 D ^2 +\frac{\lvert \mathbf{V}\rvert^4}{4 \omega ^4}},\label{nA}
	\end{equation}	 
	\begin{equation}
		\left(n_{B}\right)^2=S-\frac{\lvert \mathbf{V}\rvert^2}{2 P \omega ^2}-\frac{1}{ P}\sqrt{P^2 D ^2 +\frac{\lvert \mathbf{V}\rvert^4}{4 \omega ^4}}.\label{nB}
\end{equation}

From \eqref{SLmatriz2.3}, mixed elliptical polarizations are evaluated for the propagating modes,
\begin{subequations}
\label{polarization-parameters-for-Faraday-configuration-with-background-orthogonal-1}
\begin{equation}
		n_{A,B} \quad \rightarrow \quad \mathbf{E}_{A,B}=
		C\begin{bmatrix}
			-\zeta\\
			1\\
			-i \lvert \mathbf{V}\rvert\left(n_{A,B}^{2}-S\right) /(\omega \bar{\kappa} P)
		\end{bmatrix},\label{nABmodes}
	\end{equation}%
where
	\begin{equation}
		C=\frac{\lvert \bar{\kappa}\rvert}{\sqrt{\left(n^2-S\right)^2\left[\left(\lvert \mathbf{V}\rvert/\omega P\right)^2+\sin^2\phi\right]+D^2\cos^2\phi+\lvert \bar{\kappa}\rvert^2}},
	\end{equation}
	\begin{equation}
		\zeta=\frac{\left(  \left(  n^{2}-S\right)  ^{2}-D^{2}\right)  \cos\phi
			\sin\phi+iD\left(  n^{2}-S\right)  }{\left\vert \bar{\kappa}\right\vert ^{2}},
	\end{equation}	
	\begin{equation}
		\bar{\kappa}=\left(n^2-S\right)\cos\phi-iD\sin\phi.
\end{equation}
\end{subequations}

In \eqref{nABmodes}, one notices a longitudinal imaginary component. The transversal sector is, in general, elliptically polarized since $\zeta$ is a complex quantity. Starting from Eq. (\ref{nABmodes}) and setting $\lvert \mathbf{V}\rvert\rightarrow0$, the usual RCP and LCP modes are recovered, which is an expected correspondence.

The refractive indices given in (\ref{nA}) and (\ref{nB}) have real positive roots given by
\begin{align}
\omega_{A1,B}&=\frac{1}{6}\sqrt{8\omega_{c}^{2}\mp2\sqrt[3]{\frac{\mp2}{U}%
	}f_{1}+(\mp2)^{2/3}\sqrt[3]{U}+2\left\vert \mathbf{V}\right\vert ^{2}},    \label{frequency-omega-A1-B} \\
 \omega_{A2}&=\frac{1}{6}\sqrt{8\omega_{c}^{2}+\frac{(2)^{4/3}(-1)^{2/3}%
		f_{1}}{\sqrt[3]{U}}- f_{4}}, \label{frequency-omega-A2}
\end{align}
where
\begin{subequations}
\begin{align}
U&=11\omega_{c}^{6}+f_{2}+3\sqrt{3}\sqrt{-c^{4}\left(  5\omega_{c}^{8}%
	+f_{3}\right)  }, \\
f_{1}&=4\omega_{c}^{4}+2\omega_{c}^{2}\left\vert \mathbf{V}\right\vert
^{2}+\left\vert \mathbf{V}\right\vert ^{4}, \\
f_{2}&=-12\omega_{c}^{4}\left\vert \mathbf{V}\right\vert ^{2}+6\omega_{c}%
^{2}\left\vert \mathbf{V}\right\vert ^{4}+2\left\vert \mathbf{V}\right\vert
^{6}, \\
f_{3}&=24\omega_{c}^{6}\left\vert \mathbf{V}\right\vert ^{2}+4\omega_{c}%
^{4}\left\vert \mathbf{V}\right\vert ^{4}+12\omega_{c}^{2}\left\vert
\mathbf{V}\right\vert ^{6}+4\left\vert \mathbf{V}\right\vert ^{8}, \\
f_{4} &=\sqrt[3]{-1}\left(  2\right)  ^{2/3}\sqrt[3]{U}+2\left\vert \mathbf{V}\right\vert ^{2}, 
\end{align}
\end{subequations}
where $\omega_{A1,2}$ is associated with $n_{A}$ and $\omega_{B}$ is associated with $n_{B}$.

\subsubsection{About the index $n_{A}$ \label{secNA}}

\begin{figure}[h]		
	\centering
	\includegraphics[scale=0.72]{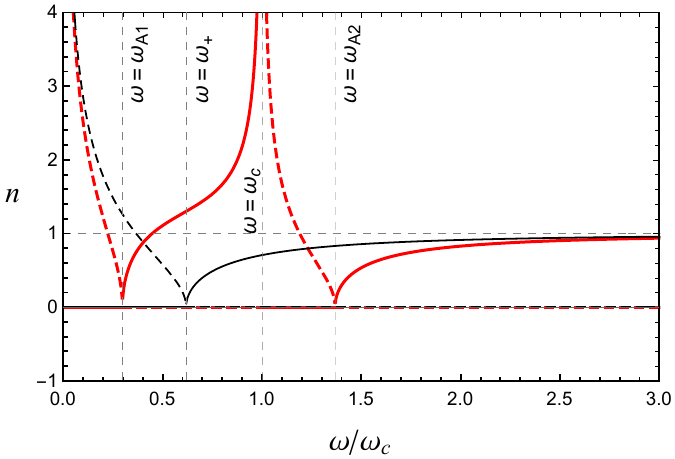}
	\caption{Refractive index  $n_{A}$ (red lines) and $n_{+}$ (black lines) of \eqref{nusual2}. The dashed (solid) lines correspond to the imaginary (real) pieces of $n_{A}$ and $n_{+}$. There is an intermediary propagation zone in the range $\omega_{A1}<\omega<\omega_{c}$ and a new absorption interval defined for $\omega_{c}<\omega<\omega_{A2}$. Here, $\omega_{c}=\omega_{p}$, $\lvert \mathbf{V}\rvert=2\omega_{p}$, and $\omega_{c}=1$~$\mathrm{rad}$~s$^{-1}$. The (solid and dashed) red curves are thicker than the (solid and dashed) black lines.}
	\label{nAfig}
\end{figure}

The refractive index $n_{A}$, given in \eqref{nA}, has two positive roots, $\omega_{A1}$ and $\omega_{A2}$, as illustrated in Fig.~\ref{nAfig}, where we observe  the following:
\begin{enumerate}
	[label=(\roman*)]
	
	\item For $\omega\rightarrow0$, the index is imaginary and tends to infinity, $n_{A}\rightarrow+i\infty$, which is the same behavior as for the usual magnetized plasma index $n_{+}$ near the origin.
	
	\item For $0<\omega<\omega_{A1}$, it holds that $\mathrm{Re}\left[n_{A}\right]= 0$, $\mathrm{Im}\left[n_{A}\right]\neq 0$, defining an absorption zone.
	
	\item  For $\omega_{A1}<\omega<\omega_{c}$, $n_{A}$ is real, $\mathrm{Re}\left[n_{A}\right]>0$ and $\mathrm{Im}\left[n_{A}\right]=0$, opening an intermediary propagation zone that does not appear in the usual case. Compare the black and red lines in Fig.~\ref{nAfig}.

	\item For $\omega\rightarrow\omega _{c}$, there is a resonance, $n_{A}\rightarrow\infty$, at the cyclotron frequency, a behavior
	that is not seen in the standard case. For $\omega_{c}<\omega<\omega_{A2}$, $n_{A}$ is imaginary, that is, $\mathrm{Im}\left[n_{A}\right]\neq0$, $\mathrm{Re}\left[n_{A}\right]=0$, and thus another absorption zone is allowed.
	
	\item For $\omega>\omega_{A2}$, there is a propagating zone, in which the index $n_{A}$ is always positive and real, with $n_{A} \rightarrow 1$ in the high-frequency limit. See Fig.~\ref{nAfig}.
	
\end{enumerate}

\subsubsection{About the index $n_{B}$ \label{secNB}}

\begin{figure}[h]		
	\centering
	\includegraphics[scale=0.72]{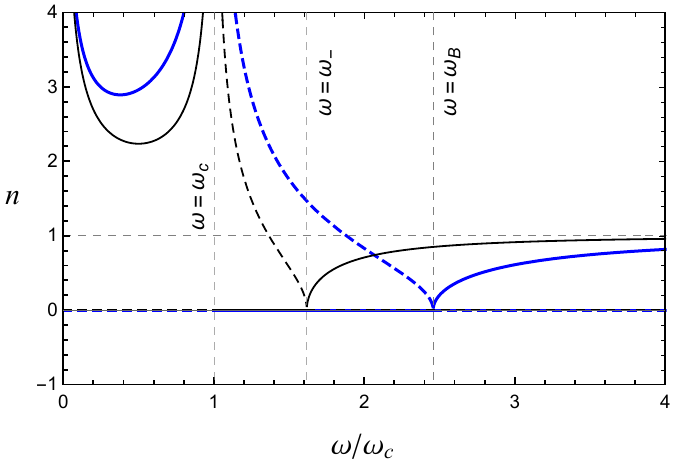}
	\caption{Refractive index $n_{B}$ (blue lines) and $n_{-}$ (black lines) of \eqref{nusual2}. The dashed (solid) lines correspond to the imaginary (real) pieces of $n_{B}$ and $n_{-}$. The intermediary absorption zone, $\omega_{c}<\omega<\omega_{B}$, has now amplified length. Here, $\omega_{c}=\omega_{p}$, $\lvert \mathbf{V}\rvert=2\omega_{p}$, and $\omega_{c}=1$~$\mathrm{rad}$~s$^{-1}$. The (solid and dashed) blue curves are thicker than the (solid and dashed) black lines. }
	\label{nBfig}
\end{figure}

The index $n_{B}$ in Eq. (\ref{nB}) has only one cutoff frequency, represented by $\omega_{B}$.  Its real and imaginary parts are illustrated in Fig.~\ref{nBfig}.

\begin{enumerate}
	[label=(\roman*)]
	
	\item In the limit $\omega\rightarrow0$, the index $n_{B}$ is real and tends to infinity, $n_{B}\rightarrow+\infty$. For $0<\omega<\omega_{c}$, the index $n_{B}$ is real, with $\mathrm{Im}[n_{B}]=0$, a behavior close to that of standard magnetized plasmas.
	
	\item For $\omega\rightarrow\omega _{c}$, $n_{B}\rightarrow\infty$,
		and there is a resonance at the cyclotron frequency. This behavior also occurs in the usual case. For $\omega_{c}<\omega<\omega_{B}$, the index is imaginary, $\mathrm{Re}[n_{B}]=0$, $\mathrm{Im}[n_{B}]\neq 0$, describing an absorption zone, that is larger than the usual one, since $\omega_{B}>\omega_{-}$. See Fig.~\ref{nBfig}.
	
	\item For $\omega>\omega_{B}$, one has a propagating zone, where $n_{B}$ is always real and positive, with $n_{B} \rightarrow1$ in the high-frequency limit. 
	
\end{enumerate}

The results of the present section cannot be compared with those of Ref.~\cite{Filipe1} since the defining condition of this section -- a chiral vector orthogonal to the magnetic field -- is not possible in the scenario of a scalar chiral factor \cite{Filipe1}.

\subsubsection{Optical effects \label{OEVOB}}

Considering the configuration of the background vector orthogonal to the magnetic field, the refractive indices (\ref{nA}) and (\ref{nB}) are not associated with circularly polarized modes [see \eqref{nABmodes}]. In this panorama, the birefringence is better characterized in terms of the phase shift per unit length, given by
	\begin{equation}
		\frac{\Delta}{d}=\frac{2\pi}{\lambda_{0}}\left(\mathrm{Re}[n_{A}]-\mathrm{Re}[n_{B}]\right),
	\end{equation}
	or explicitly
	\begin{equation}
	\label{phase-shift-Faraday-configuration-background-orthogonal-1}
		\frac{\Delta}{d}=\frac{2\pi}{\lambda_{0}}\left(\sqrt{S-\frac{\lvert \mathbf{V}\rvert^2}{2 P \omega ^2}+\Xi_{AB}}-\sqrt{S-\frac{\lvert \mathbf{V}\rvert^2}{2 P \omega ^2}-\Xi_{AB}}\right),
	\end{equation}
	where
	\begin{equation}
		\Xi_{AB}\left(\omega,\lvert \mathbf{V}\rvert\right)=\sqrt{ D ^2 +\frac{\lvert \mathbf{V}\rvert^4}{4 P^2 \omega ^4}}.
\end{equation}

In the high-frequency limit $\omega >> \left(\omega_{c},\omega_{p}\right)$, the phase shift is  

	\begin{equation}
	\frac{\Delta}{d} \simeq \, -\frac{2\pi}{\lambda_{0}}\sqrt{\frac{2\omega_{c}\omega_{p}^{2}}{\left(2 \omega ^2-\lvert \mathbf{V}\rvert^2\right)\omega}}.
\end{equation}

	As for the absorption effect for noncircularly propagating modes, one can define the difference of absorption between the two modes per unit length, written as 
	\begin{equation}
		\frac{\Delta_{Im}}{d}=\frac{2\pi}{\lambda_{0}}\left(\mathrm{Im}[n_{A}]-\mathrm{Im}[n_{B}]\right).\label{phaseshiftIM}
	\end{equation}
	For $\omega_{c}<\omega<\omega_{A2}$, the indices $n_{A}$ and $n_{B}$ are purely imaginary; see the corresponding dashed lines in Fig.~\ref{nAfig} and \ref{nBfig}. In this range, the absorption factor (\ref{phaseshiftIM}) is
	\begin{equation}
	\label{absorption-difference-Faraday-configuration-background-orthogonal-1}
		\frac{\Delta_{Im}}{d}=\frac{2\pi}{\lambda_{0}}\left[\sqrt{\frac{\lvert \mathbf{V}\rvert^2}{2 P \omega ^2}-\Xi_{AB}-S}-\sqrt{\frac{\lvert \mathbf{V}\rvert^2}{2 P \omega ^2}+\Xi_{AB}-S}\right].
	\end{equation}
	For $\omega<\omega_{A1}$, only the mode associated with $n_{A}$ is absorbed. In this case, we can write the absorption coefficient $\gamma=2\omega\mathrm{Im}[n]$, which is explicitly given by
	\begin{equation}
	\label{absorption-difference-Faraday-configuration-background-orthogonal-2}
		\gamma=2\omega\sqrt{\frac{\lvert \mathbf{V}\rvert^2}{2 P \omega ^2}-\Xi_{AB}-S}.
\end{equation}

\section{\label{section-IV}WAVE PROPAGATION ORTHOGONAL TO THE MAGNETIC FIELD}

For propagation orthogonal to the magnetic field, we can implement $\mathbf{n}=\left(n_{x},n_{y},0\right)$  in \eqref{SLmatriz2}. Furthermore, we propose a parametrization for the wave propagation in the plane orthogonal to the magnetic field in terms of the angle $\alpha$ between the propagation direction and the $x$ axis,
\begin{equation}
	\mathbf{n}=n\left(\cos\alpha,\sin\alpha,0\right).
\end{equation}
Thus, Eq. (\ref{EWE2}) now reads
	\begin{widetext}
	\begin{equation}
	\begin{bmatrix}
	n^{2}-n^{2}\cos^{2}\alpha-S & -n^{2}\cos\alpha\sin\alpha+iD+i\left(  \lvert \mathbf{V}\rvert /\omega\right)\cos\beta & -i\left(  \lvert \mathbf{V}\rvert /\omega\right)\sin\phi\sin\beta\\
	-n^{2}\cos\alpha\sin\alpha-iD-i\left(  \lvert \mathbf{V}\rvert /\omega\right)\cos\beta & {n^{2}-n^{2}\sin^{2}\alpha-S} &
	+i\left(  \lvert \mathbf{V}\rvert /\omega\right)\cos\phi\sin\beta\\
	+i\left(  \lvert \mathbf{V}\rvert /\omega\right)\sin\phi\sin\beta & -i\left(  \lvert \mathbf{V}\rvert /\omega\right)\cos\phi\sin\beta & n^{2}-P
	\end{bmatrix}%
	\begin{bmatrix}
	\delta E_{x}\\
	\delta E_{y}\\
	\delta E_{z}%
	\end{bmatrix}
	=0. \label{SLmatriz3}
	\end{equation}
\end{widetext}
The null determinant condition provides the following dispersion relation:
\begin{equation}
 \left(n^2-P\right) \left[D ^2+S \left(n^2-S\right)\right)]-\left(  \lvert \mathbf{V}\rvert /2\omega\right)^2\Theta=0, \label{det.ort}
\end{equation} 
where
\begin{align}
\Theta&=\left[2 (P+S)-3 n^2\right]+8 D  \omega   \left(P-n^2\right)\cos\beta/\lvert \mathbf{V} \rvert \nonumber \\
&-	 \left(n^2-2 P+2 S\right)\cos (2 \beta )+2 n^2  \sin ^2(\beta ) \cos (2 (\alpha -\phi )).
\label{THETA}
\end{align}

\subsection{\label{BGVORT1}Chiral vector parallel to the magnetic field}

A chiral vector parallel to the magnetic field, $\mathbf{V}\parallel \mathbf{B_{0}}$, implies $\beta=0$, whose replacement in \eqref{det.ort} yields the index $n_{T}^{2}$, 
	\begin{equation}  
		n_{T}^{2}=1-\frac{\omega_{p}^{2}}{\omega^{2}},
		 \label{nT2B}
	\end{equation}
which is also given in \eqref{nusualort1}, and is associated with the same usual linear transversal mode.

It also provides a modified refractive index,
\begin{equation}
n_{ \chi}^{2}=\frac{S^2-D ^2}{S}-\frac{\lvert \mathbf{V}\rvert^2+2 D  \lvert \mathbf{V}\rvert \omega }{S \omega ^2},\label{nparallel}
\end{equation}
associated with the elliptical propagating mode,
	\begin{equation}
	n_{ \chi} \quad \rightarrow \quad \mathbf{E}=%
	C_{\chi}\begin{bmatrix}
	-\zeta_{\chi}\\
	1
	\\
	0
	\end{bmatrix},
	\end{equation}
	where
	\begin{subequations}
	\label{polarization-paramenters-Voigt-configuration-background-parallel-1}
	\begin{align}
	C_{\chi} &=\frac{1}{\sqrt{\lvert\zeta_{\chi}\rvert^2+1}}, \\
	\zeta_{\chi} &=\frac{i\left[D+\left(  \left\vert \mathbf{V}\right\vert
		/\omega\right)\right] -n^{2}\cos\alpha\sin\alpha}{n^{2}\sin^{2}\alpha-S}.
	\end{align}
	\end{subequations}
The refractive index $n_{ \chi}$ has three cutoff frequencies, $\omega_{R}^{\pm}$ and $\omega_{L}$, given in Eqs. (\ref{omegaR}) and (\ref{omegaL}).

\subsubsection{About the index $n_{ \chi}$}

\begin{figure}[h]		
	\centering
	\includegraphics[scale=0.72]{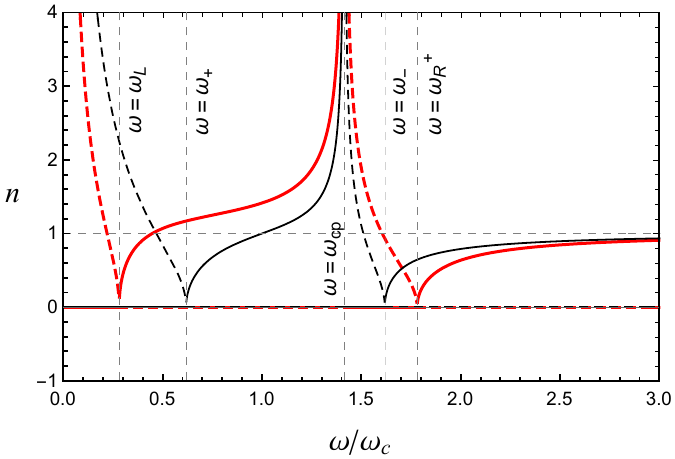}
	\caption{Red line: index $n_{ \chi}$ under the condition (\ref{omegaLcond}). Black line: index $n_{O}$. Dashed (solid) lines represent the imaginary (real) pieces of $n_{ \chi}$ and $n_{O}$. Note that the chiral vector factor shortens the first absorption zone and slightly augments the second one. Here we use: $\omega_{c}=\omega_{p}$, $\lvert \mathbf{V}\rvert=0.5\omega_{p}$, and $\omega_{c}=1$~$\mathrm{rad}$~s$^{-1}$. The (solid and dashed) red curves are thicker than the (solid and dashed) black lines.}
	\label{nort2}
\end{figure}

The refractive index $n_{ \chi}$ has the refractive index $n_{O}$ as the conventional cold plasma counterpart, given in (\ref{nusualort2}), sharing with it the same resonance frequency
\begin{equation}
	\omega_{cp}=\sqrt{\omega_{c}^{2}+\omega_{p}^{2}} .
\end{equation}
Under the condition (\ref{omegaLcond}), $n_{ \chi}$ shows two cutoff frequencies, $\omega_{R}^{+}$ and $\omega_{L}$, which are the same as those of Eqs.~(\ref{omegaR}) and (\ref{omegaL}), respectively. These frequencies are marked in Fig.~\ref{nort2}. Moreover, we point out the following:
	
\begin{enumerate}
	[label=(\roman*)]
	
	\item 	For $0<\omega<\omega _{L}$, $n_{ \chi}$ is imaginary, corresponding to an absorption zone smaller than the usual one, $0<\omega<\omega _{+}$, since $\omega_{L}<\omega _{+}$. See the black dashed line in Fig.~\ref{nort2}.
	
	\item For $\omega _{L}<\omega<\omega_{cp}$, $\mathrm{Re}[n_{ \chi}]\ne0$ and $\mathrm{Im}[n_{ \chi}]=0$, defining a propagation zone larger than the standard-case one ($\omega _{+}<\omega<\omega_{cp}$).
	
	\item For $\omega\rightarrow\omega _{cp}$,  there is a resonance, $n_{ \chi}\rightarrow+\infty$, which is the same behavior as in the usual case. For $\omega _{cp}<\omega<\omega_{R}^{+}$, $\mathrm{Re}[n_{ \chi}]=0$ and $\mathrm{Im}[n_{ \chi}]\ne0$, and another absorption zone is allowed.
	
	\item For $\omega>\omega_{R}^{+}$, the quantity $n_{ \chi}$ is always positive, corresponding to a propagation zone, which in the usual case begins at $\omega=\omega_{-}$. 
	
\end{enumerate}

\begin{figure}		
	\centering
	\includegraphics[scale=0.72]{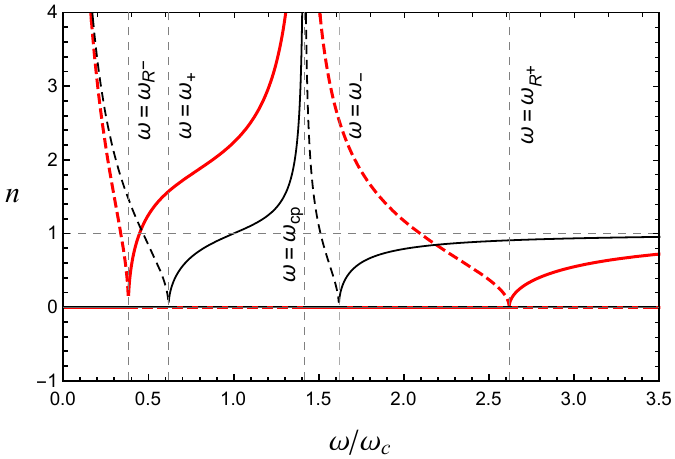}
	\caption{Red line: plot of the index $n_{ \chi}$ under the condition (\ref{omegaRcond}). Black line: plot of the index $n_{O}$. Dashed (solid) lines represent the imaginary (real) pieces of $n_{ \chi}$ and $n_{O}$. The chiral vector reduces the first absorption frequency window and enhances the second one. Here we use: $\omega_{c}=\omega_{p}$, $\lvert \mathbf{V}\rvert=2\omega_{p}$, and $\omega_{c}=1$~$\mathrm{rad}$~s$^{-1}$. The (solid and dashed) red curves are thicker than the (solid and dashed) black lines.}
	\label{nort}
\end{figure}

Under the condition (\ref{omegaRcond}), $n_{ \chi}$ has two roots, $\omega_{R}^{\pm}$, and similar characteristics to those described above, as shown in Fig.~\ref{nort}.

\subsubsection{Optical effects \label{secRP2}}

For the configuration of a background vector parallel to the magnetic field, the propagating modes obtained were described by elliptical and linear polarized vectors, associated with the refractive indices $n_{T}$ and $n_{ \chi}$, respectively. In this case, the birefringence is evaluated by employing the phase shift per unit length,
	\begin{equation}
	\label{phase-shift-equation-1}
		\frac{\Delta}{d}=\frac{2\pi}{\lambda_{0}}\left(\mathrm{Re}[n_{T}]-\mathrm{Re}[n_{ \chi}]\right),
	\end{equation}
which for the indices (\ref{nT2B}) and (\ref{nparallel}) reads
	\begin{equation}
	\label{phase-shift-Voigt-configuration-backgroud-parallel-1}
		\frac{\Delta}{d}=\frac{2\pi}{\lambda_{0}}\left(\sqrt{P}-\sqrt{\frac{S^2-D ^2}{S}-\frac{\lvert \mathbf{V}\rvert^2+2 D  \lvert \mathbf{V}\rvert \omega }{S \omega ^2}}\right).
	\end{equation}
	In the limit $\omega>>(\omega_{p},\omega_{c})$, using the parameters $S$, $D$, and $P$ given in (\ref{def}), such a phase shift reduces to 
	\begin{equation}
		\frac{\Delta}{d}=\frac{2\pi}{\lambda_{0}}\left(1-\sqrt{\frac{\omega ^2-\lvert \mathbf{V}\rvert^2}{\omega ^2}}\right). \label{PSnTnParallel}
	\end{equation}
	In the usual case, where $\Delta/d\propto(n_{T}-n_{O})$, the phase shift is null in the high-frequency limit, a result recovered for $\lvert \mathbf{V}\rvert\rightarrow0$ in (\ref{PSnTnParallel}). Thus, the chiral vector is responsible for an unusual dispersive birefringence in the high-frequency domain.

Concerning the absorption for noncircularly propagating modes, the difference in absorption between the two modes per unit length,
	\begin{equation}
		\frac{\Delta_{Im}}{d}=\frac{2\pi}{\lambda_{0}}\left(\mathrm{Im}[n_{T}]-\mathrm{Im}[n_{ \chi}]\right),
	\end{equation}
yields
	\begin{equation}
	\label{absorption-difference-Voigt-configuration-background-parallel-1}
		\frac{\Delta_{Im}}{d}=\frac{2\pi}{\lambda_{0}}\left(\sqrt{P}-\sqrt{\frac{\lvert \mathbf{V}\rvert^2+2 D  \lvert \mathbf{V}\rvert \omega }{S \omega ^2}-\frac{S^2-D ^2}{S}}\right),
	\end{equation}
for the indices in Eqs.~(\ref{nT2B}) and (\ref{nparallel}). It is non-null for $0<\omega<\omega_{L,R^{-}}$ and $\omega_{cp}<\omega<\omega_{R^{+}}$, under the condition (\ref{omegaRcond}) or (\ref{omegaLcond}), as shown in Figs.~\ref{nort2} and \ref{nort}, respectively.

\subsection{\label{BGVORT2}Background vector orthogonal to the magnetic field}

Considering the chiral vector orthogonal to the magnetic field, $\mathbf{V}\perp \mathbf{B}_{0}$, we take $\beta \rightarrow \pi/2$ in \eqref{det.ort}, yielding two refractive indices
	\begin{equation}
		(n^{2}_{\Upsilon})_{\pm}=\frac{ S (P+S)-D ^2}{2S}-\frac{\lvert \mathbf{V}\rvert^2 \sin ^2(\alpha -\phi )}{2S\omega^{2}} \pm \frac{\eta}{4S\omega^2}\label{indicesB},
	\end{equation}
where
\begin{subequations}
\label{refractive-parameters-Voigt-configuration-background-orthogonal-1}
\begin{align}
\eta &= \sqrt{\Lambda^{2}-16S\omega^2\left(  P\omega^{2}\left(  S^{2}%
	-D^{2}\right)  -S\lvert \mathbf{V}\rvert^{2}\right)  }, \\
\Lambda &= 2\left(  D^{2}-S^{2}-PS\right)  \omega^{2}-\lvert \mathbf{V}\rvert^{2}%
	\cos\left(  2\left(  \alpha-\phi\right)  \right)  +\lvert \mathbf{V}\rvert^{2}.
\end{align}
\end{subequations}
The indices $(n^{2}_{\Upsilon})_{\pm}$ are related to the following electromagnetic modes:
\begin{align}
	(n_{\Upsilon})_{\pm} \quad \rightarrow \quad \mathbf{E_{\pm}} =
	C_{\Upsilon} \begin{bmatrix}
	-\zeta_{\Upsilon}\\
     1 \\
	\frac{\lvert\mathbf{V} \rvert\left(
		\cos\phi+i\zeta_{\Upsilon}\sin\phi\right)  }{\omega\left(  n_{\pm}^{2}-P\right) }
	\end{bmatrix},
	\end{align}	
	where $C_{\Upsilon}$ is a normalization constant and
	\begin{subequations}
	\label{polarization-parameters-Voigt-configuration-background-orthogonal-1}
	\begin{align}
	\zeta_{\Upsilon}  &=    \frac{\zeta_{*}+iD\left(  n^{2}\sin^{2}%
		\alpha-S\right)  \left(  \cos^{2}\phi-\sin^{2}\phi\right) }{\left\vert
		\bar{\lambda}\right\vert ^{2}},  
	\end{align}
with	
	\begin{align}
	\zeta_{*} &=\gamma  +D^{2}\cos
	^{2}\phi-n^{2}\cos\alpha\sin\alpha\left(  n^{2}\sin^{2}\alpha-S\right), \\
	\gamma&= \left[ \left(  n^{2}\sin^{2}\alpha-S\right)  ^{2}+\left(
	n^{2}\cos\alpha\sin\alpha\right)  ^{2}\right]\sin\phi\cos\phi, \\
\bar{\lambda}  &=\left(  n^{2}\sin^{2}\alpha-S\right)  \cos\phi-\left(
n^{2}\cos\alpha\sin\alpha+iD\right)  \sin\phi.
\end{align}
\end{subequations}

The refractive indices in (\ref{indicesB}) have real and positive roots $\omega_{\perp1,2}$ associated with $\left(n_{\Upsilon}\right)_{+}$ and $\omega_{\perp+}$ associated with $\left(n_{\Upsilon}\right)_{-}$. These frequencies are not presented here as they are very extensive and intricate solutions of a sixth-order equation in frequency.

In the following, some aspects of the indices $(n_{\Upsilon})_{\pm}$ will be discussed. Figures \ref{nperp} and \ref{nperp2} illustrate the general behavior of $(n_{\Upsilon})_{\pm}$ for $\alpha-\phi=0$ ($\mathbf{V}\parallel \mathbf{n}$) and $\alpha-\phi=\pi/2$ ($\mathbf{V}\perp \mathbf{n}$), shown by the red and blue lines, respectively.
	
\subsubsection{About the index $(n_{\Upsilon})_{+}$}

The refractive index $(n_{\Upsilon})_{+}$ can be compared to the index $n_{T}$, given in \eqref{nT2B}, which describes the usual transversal mode. We find that $(n_{\Upsilon})_{+}$ has two cutoff frequencies, $\omega_{\perp}$ and $\omega_{\perp2}$. The behavior of $(n_{\Upsilon})_{+}$ is illustrated in Fig.~\ref{nperp}, which shows the following features:

\begin{figure}[h]		
	\centering
	\includegraphics[scale=0.72]{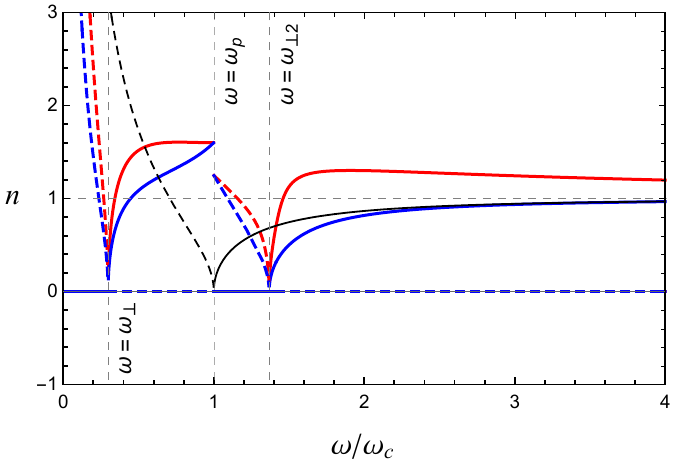}
	\caption{Red (blue) line indicates the index $(n_{\Upsilon})_{+}$ for $\alpha-\phi=0$ ($\alpha-\phi=\pi/2$). The black line illustrates $n_{T}$. Dashed (solid) lines represent the imaginary (real) pieces of $(n_{\Upsilon})_{+}$ and $n_{T}$.  The chiral factor allows a new intermediary propagating zone for $\omega_{\perp}<\omega<\omega_{p}$, which is followed by a new absorption zone for $\omega_{p}<\omega<\omega_{\perp2}$. Here we use: $\omega_{c}=\omega_{p}$, $\lvert\mathbf{V}\rvert=2\omega_{p}$, and $\omega_{c}=1$~$\mathrm{rad}$~s$^{-1}$. The black line is the thinnest one (solid and
		dashed parts). The blue line is lower than the red curve (solid
		and dashed pieces).}
	\label{nperp}
\end{figure}

\begin{enumerate}
	[label=(\roman*)]
	
	\item For $0<\omega<\omega _{\perp}$, $(n_{\Upsilon})_{+}$ is imaginary, corresponding to an absorption zone. 
	
	\item For $\omega_{\perp}<\omega<\omega_{p}$, there is an attenuation-free propagation zone where $\mathrm{Re}[(n_{\Upsilon})_{+}]\ne0$ and $\mathrm{Im}[(n_{\Upsilon})_{+}]=0$. In the standard case, there is an absorption zone in this range.
	
	\item For $\omega\rightarrow\omega _{p}$, $(n_{\Upsilon})_{+}$ has an unusual discontinuity, as shown in Fig.~\ref{nperp}. For $\omega_{p}<\omega<\omega_{\perp2}$, the index $(n_{\Upsilon})_{+}$ is imaginary and there is an absorption zone. This aspect contrasts with the usual case, where $n_{T}$ is always real for $\omega>\omega_{p}$.
	
	\item For $\omega>\omega_{\perp2}$, $(n_{\Upsilon})_{+}$ is real, yielding an attenuation-free propagation zone.

\end{enumerate}

\subsubsection{About the index $(n_{\Upsilon})_{-}$}

\begin{figure}		
		\centering
		\includegraphics[scale=0.72]{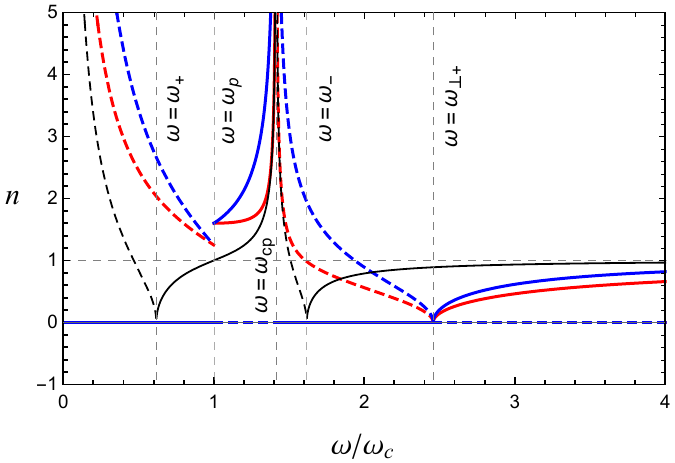}
		\caption{Red (blue) line: plot of the index  $(n_{\Upsilon})_{-}$ for $\alpha-\phi=0$ ($\alpha-\phi=\pi/2$). Black line: plot of the index $n_{O}$. Dashed (solid) lines represent the imaginary (real) pieces of  $(n_{\Upsilon})_{-}$ and $n_{O}$. The chiral factor enhances the length of the two absorption zones. Here, we have used: $\omega_{c}=\omega_{p}$, $\lvert\mathbf{V}\rvert=2\omega_{p}$, and $\omega_{c}=1$~$\mathrm{rad}$~$s^{-1}$.   The black
			line is the thinnest one (solid and dashed parts). The red line is
			lower than the blue curve (solid and dashed pieces). }
		\label{nperp2}
	\end{figure}

The index $(n_{\Upsilon})_{-}$ is a modification of the index $n_{O}$, given in \eqref{nusualort2}, associated with the usual extraordinary mode. The former has a cutoff frequency at $\omega_{\perp+}$, as shown in Fig.~\ref{nperp2}. Some aspects of $(n_{\Upsilon})_{-}$ are summarized below.

\begin{enumerate}
	[label=(\roman*)]
	
	\item For $0<\omega<\omega _{p}$, there is an absorption zone where $\mathrm{Re}[(n_{\Upsilon})_{-}]=0$ and $\mathrm{Im}[(n_{\Upsilon})_{-}]\ne0$.  For $\omega\rightarrow\omega _{p}$, $(n_{\Upsilon})_{-}$ has a discontinuity, as noticed in Fig.~\ref{nperp2}. For $\omega _{p}<\omega<\omega_{cp}$, the index $(n_{\Upsilon})_{-}$ is real, corresponding to a propagation window. As $\omega_{p}>\omega_{+}$, the first absorption zone is enlarged while the first propagating window is shortened.

		\item For $\omega\rightarrow\omega _{cp}$, $(n_{\Upsilon})_{-}\rightarrow \infty$, which is the same behavior as in the usual case, as indicated in Fig.~\ref{nperp2}. For $\omega_{cp}<\omega<\omega_{\perp+}$, the index $(n_{\Upsilon})_{-}$ is purely imaginary and the associated mode is absorbed. As $\omega_{\perp+}>\omega_{-}$, this second absorption zone is also enlarged in comparison with the usual-case one. 
	
	\item For $\omega>\omega_{\perp+}$, the quantity $(n_{\Upsilon})_{-}$ is always real, corresponding to an attenuation-free propagation zone. In the standard case, the propagation zone occurs for $\omega>\omega_{-}$.

\end{enumerate}

\subsubsection{Optical effects \label{secRP2B}}

For this configuration, there are two elliptical propagating modes associated with the refractive indices $(n_{\Upsilon})_{\pm}$, given in \eqref{indicesB}. Thus, the birefringence is measured in terms of the phase shift per unit length,
	\begin{equation}
	\frac{\Delta}{d}=\frac{2\pi}{\lambda_{0}}\left(
	(n_{\Upsilon})_{+}-(n_{\Upsilon})_{-}\right).
	\end{equation}
	Using the indices (\ref{indicesB}), the latter becomes
	\begin{equation}
	\frac{\Delta}{d}=\frac{2\pi}{\lambda_{0}}\left(\sqrt{\Pi-\Xi_{\perp-}}-\sqrt{\Pi-\Xi_{\perp+}}\right)\label{PS2},
	\end{equation}
	where
	\begin{subequations}
	\label{birefringence-parameters-Voigt-configuration-background-orthogonal-1}
	\begin{align}
	\Pi &=\frac{ S (P+S)-D ^2}{2S}, \\
	\Xi_{\perp\pm}\left(\omega,\lvert \mathbf{V}\rvert\right) &=\frac{\lvert \mathbf{V}\rvert^2 \sin ^2(\alpha -\phi )}{2S\omega^{2}} \pm \eta/4.
	\end{align}
	\end{subequations}
In the high-frequency limit, where $\omega>>\left(\omega_{c},\omega_{p}\right)$, Eq. (\ref{PS2}) becomes
\begin{subequations}
\label{birefringence-parameters-Voigt-configuration-background-orthogonal-2}
	\begin{align}
	\frac{\Delta}{d}=\frac{\pi}{\lambda_{0}}\left(\sqrt{4+\lvert \mathbf{V}\rvert\xi_{+}/\omega^{2}}-\sqrt{4+\lvert \mathbf{V}\rvert\xi_{-}/\omega^{2}}\right)\label{PS2.2},
\end{align}
with
	\begin{align}
	\xi_{\pm} &= \pm\sqrt{\lvert \mathbf{V}\rvert^2+8 \omega ^2+ \tilde{\gamma} \cos (2 (\alpha -\phi ))} \nonumber \\
	&\phantom{=}  -2 \lvert \mathbf{V}\rvert^2 \sin^2(\alpha-\phi), \\
	\tilde{\gamma} &=\left(8 \omega ^2-2 \lvert \mathbf{V}\rvert^2\right)+\lvert \mathbf{V}\rvert^2 \cos(2 (\alpha -\phi )). 
	\end{align}
	\end{subequations}
In this limit, the usual-case result, $\Delta/d=0$, is recovered for $\lvert \mathbf{V}\rvert\rightarrow0$.

Considering the lossy effect, one observes that electromagnetic modes associated with the refractive indices $(n_{\Upsilon})_{\pm}$ are absorbed for $\omega<\omega_{\perp}$ (see the dashed line in Figs.~\ref{nperp} and \ref{nperp2}). In this range, we can write the difference of absorption between the two modes per unit length, 
	\begin{equation}
	\frac{\Delta_{Im}}{d}=\frac{2\pi}{\lambda_{0}}\left(\mathrm{Im}[(n_{\Upsilon})_{+}]-\mathrm{Im}[(n_{\Upsilon})_{-}]\right),\label{PSIM2}
	\end{equation}
	or
	\begin{equation}
	\frac{\Delta_{Im}}{d}=\frac{2\pi}{\lambda_{0}}\left(\sqrt{\Xi_{\perp-}-\Pi}-\sqrt{\Xi_{\perp+}-\Pi}\right)\label{PSIM}.
	\end{equation}
	For $\omega_{cp}<\omega<\omega_{\perp+}$, only $(n_{\Upsilon})_{-}$ has a non-null imaginary piece. Then, the corresponding absorption coefficient $\gamma=2\omega\mathrm{Im}[(n_{\Upsilon})_{-}]$ in this range is
	\begin{equation}
	\label{PSIM-33}
	\gamma=2\omega\sqrt{\Xi_{\perp+}-\Pi}.
	\end{equation} 
	On the other hand, for  $\omega_{p}<\omega<\omega_{\perp2}$, only $(n_{\Upsilon})_{+}$ has an imaginary piece, which implies the following absorption coefficient:
	\begin{equation}
	\label{PSIM-34}
	\gamma=2\omega\sqrt{\Xi_{\perp-}-\Pi}.
	\end{equation} 
	In the usual case, the two electromagnetic modes are absorbed for $\omega<\omega_{+}$, since the indices $n_{T}$ and $n_{O}$ are purely imaginary in this range.

{\section{Final remarks and Perspectives \label{conclusion}}}

Electromagnetic-wave propagation and absorption in a cold magnetized plasma were analyzed in the case of the spacelike MCFJ {theory, which entails the AHE current term, $\mathbf{J}_{AH}=\mathbf{k}_{AF} \times\mathbf{E}$. Here, the background vector represents the chiral factor of the system.} Using the usual cold plasma permittivity tensor and the modified Maxwell equations, we obtained the dispersion relation and corresponding refractive indices for two main situations: (i) wave propagation along the magnetic axis (see Sec.~\ref{Wave-propagation-in-purely-time-like-background-case}) and (ii) wave propagation orthogonal to the magnetic axis (see Sec.~\ref{section-IV}). These two scenarios were examined for two possible configurations of the chiral vector: longitudinal and orthogonal to the magnetic field.

In Sec.~\ref{section-IIIA}, we discussed wave propagation along the magnetic field with the chiral vector in the same direction. The modified refractive indices $n_R$ and $n_L$ were obtained, being associated with RCP and LCP modes, respectively. Their properties were carefully examined in order to determine how the conventional propagation and absorption zones are affected. Figures ~\ref{nRfig}, \ref{nRfig2}, \ref{nLfig}, and \ref{nLfig2} display the dispersive behavior of these indices, which are also scrutinized in the plots of the dispersion relations (see Figs.~\ref{oxk_nr_1}, \ref{oxk_nr_2}, and \ref{oxk_nl_1}). The appearance of new absorption or propagation zones, as well as the length increase or reduction of these zones, are the main effects induced by the chiral vector. In contrast with the cold plasma under a scalar chiral factor \cite{Filipe1}, the present plasma model does not manifest negative refraction. 

In the very-low-frequency regime, the propagating modes were analyzed and there occurs the possibility of propagating RCP (\eqref{helicons-15}) or LCP helicons (\eqref{helicons-16}), another effect stemming from the chiral vector. However, only one of them can propagate for each choice of the chiral vector magnitude. This is an additional point of distinction in comparison with the cold plasma with a scalar chiral factor of Ref. \cite{Filipe1}, where both RCP and LCP helicons could propagate simultaneously. 

The circular birefringence in Sec.~\ref{section-IIIA} was evaluated in terms of the RP for the refractive indices $n_{R}$ and $n_{L}$, under the two conditions for the magnitude of the chiral vector, $\lvert \mathbf{V}\rvert >\omega_{p}^{2}/\omega_{c}$ and $\lvert \mathbf{V}\rvert <\omega_{p}^{2}/\omega_{c}$. The corresponding RPs are depicted in Figs.~\ref{rptl1} and \ref{rpsl2}, respectively,  being the
former endowed with sign reversion. Such an inversion occurs in scenarios of rotating plasmas \cite{Gueroult}, where the RP changes sign and decays as $1/\omega^2$ for high frequencies. It also occurs in the MCFJ chiral plasma with a timelike component, $V_0$ \cite{Filipe1}, where the RP reverses and tends to an asymptotic value, $-V_0$. In the present case, an analogous behavior occurs: the RP reverses and tends to the asymptotic value $-|V|/2$. Therefore, it is worth discussing the possibility of using the RP to characterize chiral {plasmas} described by the effective MCFJ electrodynamics (concerning the scalar or vector chiral factor). In this sense, one can compare the RP of Fig.~\ref{rptl1} with that of Fig. 14 of Ref. \cite{Filipe1} and note a substantial similarity: both are endowed with sign reversal and asymptotic negative values. The main difference between them takes place near the origin when the former RP tends to zero.  Concerning the RP depicted in Fig.~\ref{rpsl2}, the behavior is similar to those of Figs. 15 and 16 of Ref.~\cite{Filipe1} in the range $0<\omega<\omega_{c}$, but different for $\omega>\omega_{c}$, where the latter ones become positive due to the negative refraction, while the present RP is negative (see Fig.~\ref{rpsl2}). Thus, we {reinforce} that the RP behavior may constitute a route to distinguish between the MCFJ cold plasmas with scalar or vector chiral factors, {that is, cold plasmas with a magnetic current (CME) and the AHE.} As for the absorption zones, the coefficient of circular dichroism reveals a behavior analogous to that of the conventional cold plasma under the condition (\ref{omegaLcond}). 

In Sec.~\ref{BGVORT}, we considered the case of propagation along the magnetic field and the chiral vector orthogonal to it. The refractive indices were obtained and their properties were examined. The associated modes have mixed transversal and longitudinal components, with elliptical polarization in the transversal sector. Thus, the birefringence and the dichroism were measured in terms of phase shift coefficients per unit length.

The general dispersive behavior of the refractive indices obtained in this case is represented by Figs.~\ref{nAfig} and \ref{nBfig}. The zones of attenuation-free propagation and absorption are defined by several characteristic frequencies, determined by Eqs.~ (\ref{frequency-omega-A1-B}) and (\ref{frequency-omega-A2}). In comparison with the usual cold plasma scenario, some differences are noted. The dispersive refractive index of Fig.~\ref{nAfig} presents two absorption zones with a propagation regime between them. For the case depicted in Fig.~\ref{nBfig}, the absorption zone is increased by $\Delta \omega = \omega_{B}- \omega_{-}$ in relation to the usual case.

The scenario of propagation orthogonal to the magnetic field was addressed in Sec.~\ref{section-IV}, also considering the cases with a chiral vector parallel and orthogonal to the magnetic field. Besides the usual transversal mode of \eqref{nT2B}, in Sec.~\ref{BGVORT1} we obtained a second refractive index associated with a general elliptically polarized propagating mode. Its dispersive behavior under the condition (\ref{omegaLcond}) is represented in Fig.~\ref{nort2}, revealing that the chiral vector narrows the first absorption zone and slightly increases the second one, in comparison with the usual cold plasma. For condition (\ref{omegaRcond}), the chiral vector shortens the first window of absorption and greatly enhances the second one; see Fig.~\ref{nort}. The birefringence and absorption effects were evaluated in terms of the phase shift of \eqref{phase-shift-Voigt-configuration-backgroud-parallel-1} and the coefficient of \eqref{absorption-difference-Voigt-configuration-background-parallel-1}, respectively.

In Sec.~\ref{BGVORT2}, the case of the chiral vector orthogonal to the magnetic field was discussed. The intricate dispersive behaviors of the refractive indices obtained in this case are represented in Figs.~\ref{nperp} and \ref{nperp2}. Compared to the standard cold plasma, we note that in Fig.~\ref{nperp} a new absorption zone appears between the two propagation windows, while the chiral vector decreases the first one. In Fig.~\ref{nperp2}, the two absorption zones are enlarged compared to the corresponding zones of the usual cold plasma.

 General properties of three distinct cold plasma electro
dynamics, namely, (i) standard cold plasma, (ii) cold
plasma with magnetic current, (iii) cold plasma with
anomalous Hall current (the present article), are sumarized
in Tables.~\ref{tab:comparison-between-all-scenarios-2} and \ref{tab:comparison-between-all-scenarios-3} for Faraday and Voigt configurations, respectively, {comparing aspects of the propagating modes, birefringence, RP inversion, absorption, and helicon modes. }

{It is important to state that the present investigation has led to plasma solutions in a static axion scenario, $\partial_{t} \theta=0$, with a spatially dependent axion field,
\begin{equation}
\theta = \theta_{0}r,
\end{equation}	
such that $\mathbf{\nabla}\theta=cte$, similar to the one considered to examine the axionic Casimir-like effect in Ref.~\cite{Brevik2}. An interesting future perspective consists in examining plasma modes in the context of the axion Lagrangian, 
\begin{align}
	\mathrm{{\mathcal{L}}}=-\frac{1}{4}G^{\mu\nu}F_{\mu\nu}+g\theta (\mathbf{E}\cdot \mathbf{B)},\label{LaxionP}
\end{align}
and the corresponding equations of motion, 
\begin{align}
	\nabla\cdot\mathbf{D}  &=J^{0}-g\mathbf{\nabla}\theta 
	\cdot\mathbf{B}  ,\label{Coulomb1B}\\
	\nabla\times\mathbf{H} -\frac{\partial\mathbf{D}
	}{\partial t} &=\mathbf{J}  -  g(\partial_{t}\theta)    \mathbf{B}+g\mathbf{\nabla}\theta \times
	\mathbf{E} ,\label{Amp1B} 
\end{align}
in a time-dependent scenario.  For an oscillating axion field $\theta(t) = \theta_{0}\exp(i\omega_{a}t)$, considering a situation in which the axion field oscillates at the same frequency as the electromagnetic field, cold plasma modes may be found by employing the same framework as in the present work.  A model with an oscillating axion background was recently considered for examining Casimir forces \cite{Brevik1}, with an axion field $\theta = \theta_{0}\sin(\omega_{a}t)$. A plasma investigation in this time-dependent scenario seems to be a promising perspective. Another possibility is to examine optical properties with astrophysical interest, such as the time of arrival of radio
waves from pulsars, in the chiral plasma scenario of the present work and that of Ref.~\cite{Filipe1}. It may involve the evaluation of group velocity and time delay \cite{McDonald}, which can be accomplished in the regime of free propagation for each of the cases examined.}

\begin{widetext}

	\begin{table}[h]
		\caption{Propagation properties of cold plasmas in distinct contexts for the Faraday configuration $({\bf{k}}\parallel {\bf{B}})$. The symbol ``$-$'' means that the entry does not apply to the mentioned plasma model.}
		\begin{centering}
			\begin{tabular}{ C{2cm}  C{4.4cm}  C{4cm}  C{7cm} C{1cm} }
				\toprule \\[0.01ex]
				& \textbf{Cold plasma in usual electrodynamics} & \textbf{Cold plasmas in {MCFJ theory with $\mathbf{J}_{B}=k_{AF}^{0}\mathbf{B}$}} & \textbf{Cold plasmas in {MCFJ theory with anomalous Hall current (AHE)}}  \\[0.6ex]
				\colrule \\[0.6ex]
				Propagating modes & RCP and LCP & RCP  for $n_{R, M}$, LCP for $n_{L, E}$; {see Ref. \cite{Filipe1}}  & 
				\begin{tabular}{C{0.0cm} C{3.3cm}   C{3.3cm} } 
					\\[-5ex]
					& ${\bf{V}} \parallel {\bf{B}}$ & ${\bf{V}} \perp {\bf{B}}$  \\[0.4ex] 
					\colrule\\[0.5ex]
					&\vspace{-.2cm}LCP and RCP & \vspace{-.2cm}Elliptical
				\end{tabular}
				\\[0.6ex]
				\colrule \\[0.1ex]
				\\[-4ex]
				Birefringence & RP, $\delta=-\frac{\omega}{2}\left(\mathrm{Re}[n_{+}]-\mathrm{Re}[n_{-}]\right)$       & RP {$\delta_{LR}$ and $\delta_{ER}$; see Ref. \cite{Filipe1}}   & 
				\begin{tabular}{C{0.0cm} C{3.3cm}   C{3.3cm} } 
					\\[0.5ex]
					\\[-6ex]
					&RP (\ref{rptl}) & Phase shift (\ref{phase-shift-Faraday-configuration-background-orthogonal-1}) 
				\end{tabular}
				\\[5ex]
				\colrule \\[0.1ex]
				\\[-4ex]
				RP inversion &    No    &  Yes  & 
				\begin{tabular}{C{0.0cm} C{3.3cm}   C{3.3cm} } 
					\\[0.5ex]
					\\[-7ex]
					&Yes, under the condition (\ref{omegaLcond}) & --
				\end{tabular}  \\ [0.6ex]
				\colrule \\[0.1ex]
				\\[-4ex]
				Absorption & Yes (dichroism) & Yes (dichroism) &
				\begin{tabular}{C{0.0cm} C{3.3cm}   C{3.3cm} } 
					\\[0.5ex]
					\\[-7ex]
					& Yes, (dichroism) coefficient (\ref{dicrosl}) & Coefficients (\ref{absorption-difference-Faraday-configuration-background-orthogonal-1}) and (\ref{absorption-difference-Faraday-configuration-background-orthogonal-2}) 
				\end{tabular}  \\ [0.6ex]
				\colrule \\[0.1ex]
				Helicons & RCP & RCP and LCP, enabled by {the magnetic current}  & 
				\begin{tabular}{C{0.0cm} C{3.3cm}   C{3.3cm} } 
					\\[0.5ex]
					\\[-7ex]
					&  {RCP or LCP (nonsimultaneous);} see \eqref{helicons-Faraday-configuration-with-background-parallel-1} & --
				\end{tabular}  \\ [0.6ex]
				
				\botrule
			\end{tabular}
		\end{centering}
		\label{tab:comparison-between-all-scenarios-2}
	\end{table}

	\begin{table}[H]
		\caption{Propagation properties of cold plasmas in distinct contexts for Voigt configuration $({\bf{k}}\perp {\bf{B}})$. The symbol ``$-$'' means that the entry does not apply to the mentioned plasma model.}
		\begin{centering}
			\begin{tabular}{ C{2cm}  C{4.8cm}  C{4.5cm}  C{6cm} C{1cm} }
				\toprule \\[0.01ex]
				& \textbf{Cold plasmas in usual electrodynamics} & \textbf{Cold plasmas in {MCFJ theory with $\mathbf{J}_{B}=k_{AF}^{0}\mathbf{B}$}} & \textbf{Cold plasmas in {MCFJ theory with anomalous Hall current (AHE)}}    \\[0.6ex]
				\colrule \\[0.6ex]
				Propagating modes & \vspace{.3cm}linear for $n_{T}$ and elliptical for $n_{O}$ & \vspace{.3cm}mixed elliptical, in general & 
				\begin{tabular}{C{0.0cm} C{2.8cm}   C{2.9cm} } 
					\\[-5ex]
					& ${\bf{V}} \parallel {\bf{B}}$ & ${\bf{V}} \perp {\bf{B}}$  \\[0.4ex] 
					\colrule\\[0.5ex]
					& \vspace{-.2cm}Eliptical & \vspace{-.2cm}Elliptical
				\end{tabular}
				\\[0.6ex]
				\colrule \\[0.1ex]
				\\[-4ex]
				Birefringence &Phase shift & Phase shift & 
				\begin{tabular}{C{0.0cm} C{2.8cm}   C{2.8cm} } 
					\\[0.5ex]
					\\[-7ex]
					&Phase shift (\ref{phase-shift-Voigt-configuration-backgroud-parallel-1}) & Phase shift (\ref{PS2}) 
				\end{tabular}
				\\[0.6ex]
				\colrule \\[0.1ex]
				\\[-4ex]
				RP inversion &    --    &  --  & 
				\begin{tabular}{C{0.0cm} C{2.8cm}   C{2.8cm} } 
					\\[0.5ex]
					\\[-7ex]
					& -- & --
				\end{tabular}  \\ [0.6ex]
				\colrule \\[0.1ex]
				\\[-3.5ex]
				Absorption & Yes & Yes  &
				\begin{tabular}{C{0.0cm} C{2.4cm}   C{3.4cm} } 
					\\[-4ex]
					&  Yes, given by Eq. (\ref{absorption-difference-Voigt-configuration-background-parallel-1}) 
					& Yes, given by Eqs.~(\ref{PSIM}, \ref{PSIM-33}, \ref{PSIM-34})
				\end{tabular}  \\ [3ex]
				\\ [-4ex]
				\colrule \\[0.1ex]
					\\ [-4ex]
				Helicons & -- & -- & 
				\begin{tabular}{C{0.0cm} C{3.3cm}   C{3.3cm} } 
					\\[0.5ex]
					\\[-7ex]
					& -- & --
				\end{tabular}  \\ [0.6ex]

				\botrule
			\end{tabular}
		\end{centering}
		\label{tab:comparison-between-all-scenarios-3}
	\end{table}

\end{widetext}

	\begin{acknowledgments}
	
The authors express their gratitude to FAPEMA, CNPq, and CAPES (Brazilian research agencies) for their invaluable financial support. M.M.F. is supported by CNPq/Produtividade 311220/2019-3 and CNPq/Universal/422527/2021-1. P.D.S.S. is grateful to grant CNPq/PDJ 150584/23. Furthermore, we are indebted to CAPES/Finance Code 001 and FAPEMA/POS-GRAD-02575/21.

	\end{acknowledgments}

\end{document}